%% file: bledy_ksztaltow.tex
\documentclass[a4paper,fleqn,usenatbib]{mnras}

\usepackage[svgnames]{xcolor}

\usepackage{caption}
\usepackage{graphicx}

\usepackage{amsmath}
\usepackage{amsfonts}
\usepackage{amssymb}
\usepackage{mathrsfs}
\usepackage{morefloats}
\usepackage{tikz}
\usepackage{hyperref}
\hypersetup{colorlinks,citecolor=blue,filecolor=black,linkcolor=blue, urlcolor=blue}
\usepackage{gensymb}
\usepackage[author={GD},icon=Note,color=Yellow,open=true]{pdfcomment}
\usepackage{multirow}

\newcommand{\dd}{^{\circ}}

\title{
	Volume uncertainty assessment method of asteroid models from disk-integrated
	visual photometry.
}
\author[P. Bartczak, G. Dudzi\'{n}ski]
{
	P. Bartczak, G. Dudzi\'{n}ski
	\\
	Astronomical Observatory Institute, Faculty of Physics, Adam Mickiewicz
	University, S{\l}oneczna 36, 60-286 Pozna{\'n}, Poland
}

\newcommand{\hilight}[1]{#1}

\begin{document}
\maketitle

\label{firstpage}

\begin{abstract}
	\input{abstract.tex}

\end{abstract}


\input{main_text.tex}




\bibliography{literatura}
\addcontentsline{toc}{chapter}{\bibname}
\bibliographystyle{mnras}

\end{document}

%% file: abstract.tex
The need for more accurate asteroid models is perhaps secondary to the need to
measure their quality.  The uncertainties of models' parameters propagate to
quantities like volume or density -- the most important and informative
properties of asteroids -- affecting conclusions about their physical nature.
Our knowledge on \hilight{shapes and spins of small solar system bodies comes
mostly from visual, disk-integrated photometry.} In this work we present a
method for asteroid model uncertainty assessment based on visual photometry
(lightcurves and sparse-in-time absolute measurements) allowing the
determination of realistic volume uncertainty, as well as spin axis orientation,
rotational period and local surface features. The sensitivity analysis is
conducted by creating clones of the nominal model and accepting the ones that
fit the observations within a confidence level. The uncertainties of model
parameters are extracted from the extreme values found in the accepted clone
population.  Creation of such population of clones enables the conversion of a
deterministic asteroid model into stochastic one, and can be utilized to create
observation predictions with error bars. The method was used to assess the
uncertainties of fictitious test models and real targets, i.e. (21)~Lutetia,
(89)~Julia, (243)~Ida, (433)~Eros and (162173)~Ryugu.  We conclude that volumes,
and subsequently, densities of asteroids derived from lightcurve-based models
likely have vastly understated uncertainties, the biggest source of which is the
inability to establish the extent of the model along its spin axis.

%% file: main_text.tex
\section{Introduction}
\label{sec:intro}

Volume is one of the most important physical parameters of asteroids. When
combined with the mass estimate, it is essential in the bulk density
determination; which, in turn, allows us to peek inside asteroids and make
conclusions about their inner structure, like the macro and micro porosity,
chemical composition or material differentiation \citep{Scheeres2015}. Volume
can be the result of scaling of an asteroid \hilight{shape} model with absolute
measurements or a technique based on them, e.g.  stellar occultations (e.g.,
\cite{Durech11}), adaptive optics images (e.g., \cite{Hanus13}) or
thermophysical modelling (e.g., \cite{Usui2011}, \cite{Hanus2015},
\cite{Alilagoa2018}, \cite{Marciniak2018}). The \hilight{shape} models can vary
from spheres, 3-axial ellipsoids through convex to non-convex shapes, they can
be established via different methods and based on single \citep{Kaasalainen01II,
SAGE} or multiple data types at once \citep{Carry2010, ADAM15, Durech2017}.  The
latter results in model that explains different datasets simultaneously rather
than being simply scaled, which is a great advantage of multi-dataset inversion.
\hilight{Although limited to a small number of targets, it provides valuable and
detailed information about them \citep{Carry2008, Berthier2014, Pajuelo2018,
Vernazza2018}.}

Though we have seen tremendous advances in shape and spin modelling techniques
throughout decades, the problem of quantitative quality assessment of asteroid
models from disk-integrated photometry, without the presence of auxiliary data
to compare to, remains untackled. In case of lightcurve-only based models, the
uncertainty of volume -- when reported -- comes only from the precision of
supplemental data (e.g. time resolution of stellar occultation timings) used for
scaling the \hilight{shape} models. However, the uncertainty of the
\hilight{shape} model parameters, i.e. vertices describing the shape, spin axis
orientation or rotational phase, influence such fits as well. The shape model
uncertainty propagates to volume and, analogously, to any other property
determined from the \hilight{shape} model, affecting conclusions drawn for the
population of small bodies under scrutiny. Even when no absolute measurements
are available and size cannot be determined in physical units, dimensionless
volume and \hilight{shape} model parameters uncertainties are of much value in
models' quality assessment.

There definitely is a compelling need for a procedure of quality and uncertainty
assessment of asteroid models. Without it, we are not able to compare the models
and decide which is \textit{better} or \textit{worse}. What is more, we cannot
express our doubts about the models in numerical values, leaving us with vague
qualitative estimates only. Models' fit to the data, e.g.  expressed in $\chi^2$
of the fit, is insufficient to test their robustness. To give a simple example,
a spherical body can be modelled equally well with any spin axis orientation,
leaving these parameters totally uncertain. Similarly, some shape parameters can
also be unconstrained by observational dataset, meaning we do not have the means
to tell which parameters can be trusted and which are simply the artefacts of
the method used. For instance, if we observe a body only from one viewing angle,
its far side can either stretch far back or be concave. Flattening or thickening
of a \hilight{shape} model would again not be detected in relative photometric
lightcurves. This has a huge effect on the volume determination and currently we
are not equipped well enough to account for that.  Additionally, some
inescapable factors affect the models as well, like data precision, human error,
assumptions (e.g.  homogeneous albedo or mass distribution, principal axis
rotation), simplifications of underlying physics (e.g. using simplistic light
scattering law) and approximations, to name a few.

By far, the most abundant data type for the most numerous target sample is
disc-integrated photometry in visual bands and the majority of published
asteroid \hilight{shape} models are based solely on it. The Asteroid Lightcurve
Photometry Database \citep{LCDB2009}, as of October 2018, contains lightcurve
observations for 13578 objects. \hilight{In contrast, only hundreds of shape
models were obtained using other techniques.} Accordingly, as a first step, we
focused on creating a method for asteroid \hilight{shape} models' uncertainty
assessment based on, and in regard to, this type of data. The method is not
designed to search for new, better, global minimum, although some models with
better fit are found during the process. Due to the overwhelmingly large amount
of parameters and possible geometries it is impossible to scan the whole
parameter space.  Finding a model that fits observations is a job for a
modelling technique, whichever used. The presented method explores only the
proximity of the nominal solution and is based on creating clones of the tested
model. The main goal is to obtain relative, dimensionless volume uncertainty and
to expose parameters' indeterminacy caused by light variation to shape mapping
and incompleteness of observational dataset.

In section~\ref{sec:procedures} we take a quick look at available procedures
commonly used to evaluate the models. In section~\ref{sec:observations_limits}
we analyse the limits of volume accuracy due to observing geometries, and
information content of relative lightcurves and absolute photometry, i.e. the
ability to establish models' scale along their spin axis.  Next
(Sec.~\ref{sec:method}), we describe the method of asteroid \hilight{shape}
model uncertainty assessment, giving synthetic example in
section~\ref{sec:synthetic_example} followed by an evaluation of five real
targets in section~\ref{sec:real_examples}.

\section{Available quality assessment procedures}
\label{sec:procedures}

\hilight{
The term \textit{asteroid model}, or just \textit{model}, used hereafter denotes
a set of parameters describing the shape, spin axis orientation, rotational
period, phase of rotation for reference epoch and scattering law used to reflect
light of the surface.
}
Due to the lack of standard procedures of quality assessment regarding
asteroid models there is no publicly available information on models'
uncertainties, except for arbitrary in-house quality codes. Researchers, when
using models in their studies, must resort to experts' opinion, each probably based
on different criteria for model evaluation. There is a valid
question of subjectivity of such assessments, as well and the problem with the
choice of judges (see \cite{Uusitalo2015} and references therein). In
this section we take a quick look at the frequently used approaches to the
problem.

There is a common notion of three or four apparitions being sufficient to create
\textit{decent} models \citep{Kaasalainen2002}. The models are indeed often
judged by the number of apparitions and lightcurves used during the modelling.
There is no doubt that the quality and quantity of data used in modelling (e.g.
signal to noise ratio, number of lightcurves, number of points per lightcurve,
the distribution of apparitions entangled with spin axis orientation that define
geometries under which an asteroid was observed, etc.) have major influence on
the outcomes. However, each target is unique and poses different challenges.
For example, if the rotational axis' latitude is near $\pm 90\dd$
every lightcurve carries the same information, so neither the number of them, nor
the number of apparitions matter shape-wise. The situation presents itself totally
different for latitudes near $0\dd$. There cannot be a universal set of rules
to follow when observing an asteroid to guarantee a good model, as the
knowledge of the very physical parameters we are trying to find influences our
judgement. One is certain: the more top-quality data available, the better chances
we have of creating reliable models.

Another way of model quality assessment one might use would be transferring the
robustness measure (whichever one sees fit to use) of the method used in the
modeling onto the model itself. The reasoning here is, that if the technique
worked for some targets it should work for others as well. There are a few
asteroids for which we have complete set of observations regarding shape, size
and spin axis orientation -- the ones visited by spacecraft probes. \hilight{
One example found in literature is the comparison between the topographic model
of (951) Gaspra based on Galileo mission flyby data with lightcurve based one
\citep{KaasalainenAst3}.} Another example is the comparison of \textit{in situ}
model of (21) Lutetia with its lightcurve and adaptive optics-based model
\citep{Carry12}, although the evaluation was only possible for a half of the
body's true shape observed by ESA Rosetta spacecraft during 2010 flyby.  Also,
\cite{SAGE} compare lightcurve based model of (433) Eros to its true shape based
on data from NEAR Shoemaker probe.

In both cases model comparisons were used to validate modelling
techniques rather than to produce model parameters' uncertainties which reflect
information content of the datasets. Asteroids visited by probes constitute
tremendously valuable but tiny sample compared to the number of known targets
out there limiting considerably our ability to test methods.

Previous point brings us to another approach, which is making laboratory
measurements \citep{Barucci82, Barucci82_albedo, Ambrosio85} or creating
fictitious digital targets and trying to model them from synthetic set of
observations \hilight{\citep{Kaasalainen01I, Kaasalainen2005, SAGE}}. Still, it
would take tremendous amount of time to test a significant number of targets
with large spectrum of possible scenarios (orbits, data quality, rotational
phase and phase angle coverage, and surface characteristics defining how light
scatters) to be able to review methods in full. Modelling techniques are
becoming increasingly complex taking advantage of multiple data types and
sophisticated algorithms which hampers spotting limitations and drawbacks of the
methods even further.

An analysis of a family of solutions acquired from many modelling runs for a
given target can give some insight into the robustness of a model. An extreme,
but informative, case would be the modelling of a spherical body or a flattened
sphere. No matter which orientation of the rotational axis was chosen, the
resulting model would fit the data perfectly indicating insufficient information
content in the dataset. The family of solutions is often the basis for accepting
or rejecting a model by comparing it with the rest within some confidence level
of the fit. \hilight{A common practice, based on convex inversion method
creators' experience, is to compare the models with the $\chi^2$ up to the
smallest one enlarged by 5\% to 10\% \citep{Torppa2003}.} This approach is
hard to standardize, and does not yield results that allow for comparison
of the models of different targets or models from different modelling
techniques. Uniform scanning of parameter space near the best solution is also
not utilized here, not to mention that these models still carry all of the
method-specific problems and assumptions. The behavior or outcome of modelling
can provide additional insight and indicate the presence of a very wide global
minimum or plethora of local ones, manifesting as dissimilarity of the shapes
from family of solutions and/or large scatter of spin axis orientations giving
solutions with similar $\chi^2$.

Observational techniques other than disk-integrated photometry, even when not
used during the modelling, can be valuable for model confirmation and for
disambiguation of mirror solutions, e.g. \cite{Durech2011, Hanus2015}. Of
course, each technique has its own limitations that need to be understood and
taken into account. Especially, when a technique offers only 2D projections of 3D
shapes which, depending on the number of available projections, can seriously
limit the ability to test the models.

\section{Model scale along the spin axis}
\label{sec:observations_limits}

In this section we take a closer look at the ability to detect the extent
of the model along the spin axis due to observing geometries, and shape to
scattered light transformation, causing huge volume uncertainty. As stated
before, the vast majority of asteroid models are created from relative photometric
lightcurves alone, and those are extremely susceptible to this problem.
Demonstrating this, a 3-axial ellipsoid models established using amplitudes
method alone would have only $a/b$ ratio determined, with $b/c$ unknown and one
would needs to utilise differences in absolute magnitudes to assess $b/c$ ratio
\citep{asteroids2magnusson}. Assuming that the models are defined so that
spin- and $z$- axes are the same, the term \textit{z-scale} used hereafter will refer
to the extent of a model along the spin axis.

Although the effect of changing the z-scale is most apparent in absolute
magnitudes, it also changes the shape and amplitude of the lightcurves in some
favorable orientations, therefore one might argue that more advanced methods taking
into account all the features of the lightcurves can detect the z-scale with
success. As we will show, the observations have to be made for particular
epochs (which are target-specific) and with good photometric precision so that
the differences (rather small in general) are even detectable.

\begin{figure}
	\includegraphics[width=8.5cm]{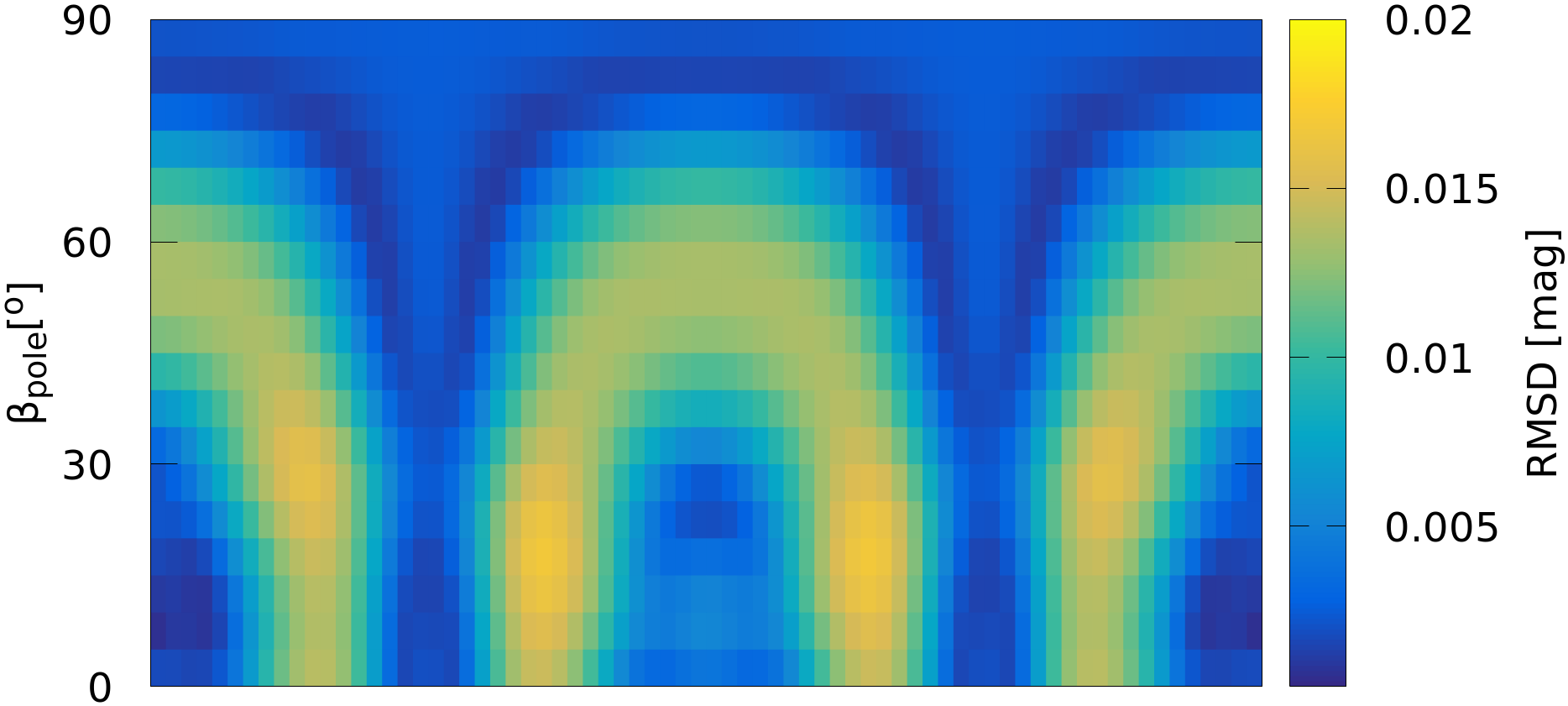}

	\vspace{0.5cm}

	\includegraphics[width=8.5cm]{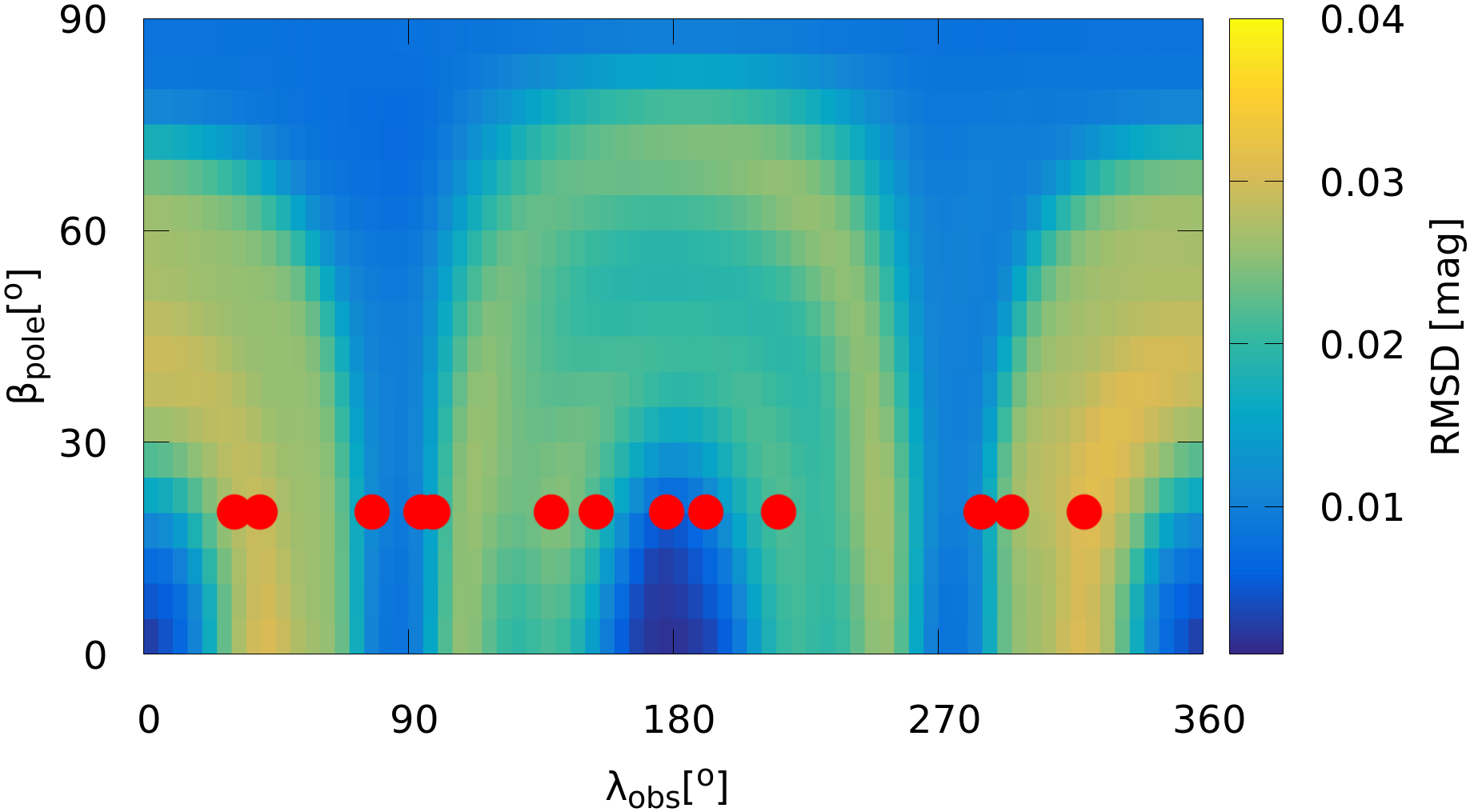}
	\caption{
Plot showing average difference of lightcurves of 3-axial ellipsoid with
$a/b=1.25$, $b/c=1$ against $b/c=2$ (top) and two (9) Metis \citep{SAGE}
models scaled in z-axis by 0.66 and 1.33 (bottom). Models were observed at
different ecliptic longitudes $\lambda_{obs}$ from $0\dd$ to $360\dd$  at
opposition with various pole latitudes $\beta_{pole}$ and Lommel-Seeliger
scattering law. Red dots in the bottom plot show observed longitudes of (9) Metis
for spin axis $\beta_{pole}=20\dd$.  See text for more details.
	}
	\label{fig:z-axis}
\end{figure}

To illustrate this phenomenon we calculated the average difference between the
lightcurves of two 3-axial ellipsoids with the same $a/b=1.25$, but different $b/c$
of $1$ and $2$ (Fig.~\ref{fig:z-axis}, top). The synthetic observations were made
for various longitudes $\lambda_{obs}$ simulating constant observations on the
orbit. The observer was put in the position of the light source observing the body always
in opposition. The orbit was coplanar with the $xy$ plane of the coordinate
system, and the Lommel-Seeliger scattering law was utilized.  We  repeated
this exercise for the model of (9)Metis from \cite{SAGE} by scaling it in
$z$-axis
by 0.66 and 1.33 (Fig.~\ref{fig:z-axis}, bottom) and using the same
observational setup as for ellipsoids.
The volume of the larger ellipsoid ($b/c=1$) is twice the volume of the
smaller one ($b/c=2$). The same holds for the two Metis models. It is important to
realize that even huge difference in volume results in very little difference in
lightcurves even for favorable geometries. The targets would have to be observed
at specific $\lambda_{obs}$ for the observer to spot the difference and the
chances to do so vary depending on the inclination of the rotational axis. Due
to the shape and orientations of the orbits of the Earth and the target, and their
orbital periods limited geometries are feasible during apparitions in reasonable
time span imposing further constrains. In our examples the biggest possible
average difference was $0.02$ mag for the ellipsoids and $0.04$ mag for the (9) Metis
models, which is very small compared to the precision of available photometric
lightcurves, which typically is $0.01$ mag. All models based solely on
relative photometry are affected by this problem, but the extent of the volume
uncertainty is case-specific for every target and set of observations.

\subsection{Absolute photometry to the rescue}
\label{sec:absolute_limit}

In theory, absolute observations contain information on the extent of an
asteroid along the spin axis. In the simple case of 3-axial ellipsoid, the
absolute flux $\Phi$ changes as a function of aspect angle $\xi$ -- the angle
between the rotation axis and asteroid-observer direction. The more
$\beta_{pole}$ approaches $0\dd$ the more evident the effect is, as $\xi$
variability increases. The ratio of the model's semi-axes $b/c$ can be found
using magnitudes method, when observations show the difference in brightness
depending on the aspect $\xi$, i.e. the change in observed projections with
exposed $a$ and $b$ axes in one apparition and $a$ and $c$ axes in the other.
This method's usefulness faints as observed aspect angles' range decreases and
fails completely for $\beta_{pole}$ approaching $\pm90\dd$ on orbits coplanar
with the ecliptic.

Adopting the formalism used by \cite{Zappala81} we can draw some conclusions
about the whole population of asteroids and our ability to determine their
volumes.  Simple example from previous paragraph hints that this ability depends
on available aspect angles under which target can be observed. Assuming a
3-axial ellipsoid described by semi-axes $a$, $b$ and $c$ and with semi-axis $c$
equal to $1$ by definition, the difference in magnitude between two observations
taken at aspects $\xi=90\dd$ (equator on) and $\xi$ is

\begin{equation}
		\Delta M = M(90\dd) - M(\xi) = 2.5\log d,
	\label{eq:DM}
\end{equation}
where
\begin{equation}
		d=\sqrt{b^2 \cos^2(\xi) + \sin^2(\xi)}
\end{equation}
is a projection area of the ellipsoid for a given aspect $\xi$.

Fig.~\ref{fig:bees} illustrates this dependency for different values of
semi-axis $b$. Observation at aspect $\xi=90\dd$ can be in principle obtained
for every asteroid, regardless of the spin axis latitude $\beta_{pole}$ if we
neglect the asteroid orbit's longitude of ascending node $\Omega$ and spin axis
longitude $\lambda_{pole}$. The red dots in Fig.~\ref{fig:bees} show a
difference in magnitudes from photometric measurements with magnitude precision
(represented by error bars) of $\delta m = 0.1$, which is a typical precision of
majority of available sparse-in-time absolute photometry from automated surveys
\citep{Hanus2011}. The non-zero precision in magnitude creates ambiguity in
calculated $b$ parameter, i.e. the true $b$ can be confused  with different one
which leads to ambiguity in volume. In case of 3-axial ellipsoid the change
in $b$ causes the same change in the volume of the body. If we assume that the
$b$ parameter of an ellipsoid can be derived from the difference $\Delta M$, we
are able to estimate, roughly, the uncertainty of the volume for a given target
observed at $\xi=90\dd$ and minimal achievable aspect angle $\xi_{\text{min}}$.
This can be expressed by the equality:

\begin{figure}
	\includegraphics[width=8cm]{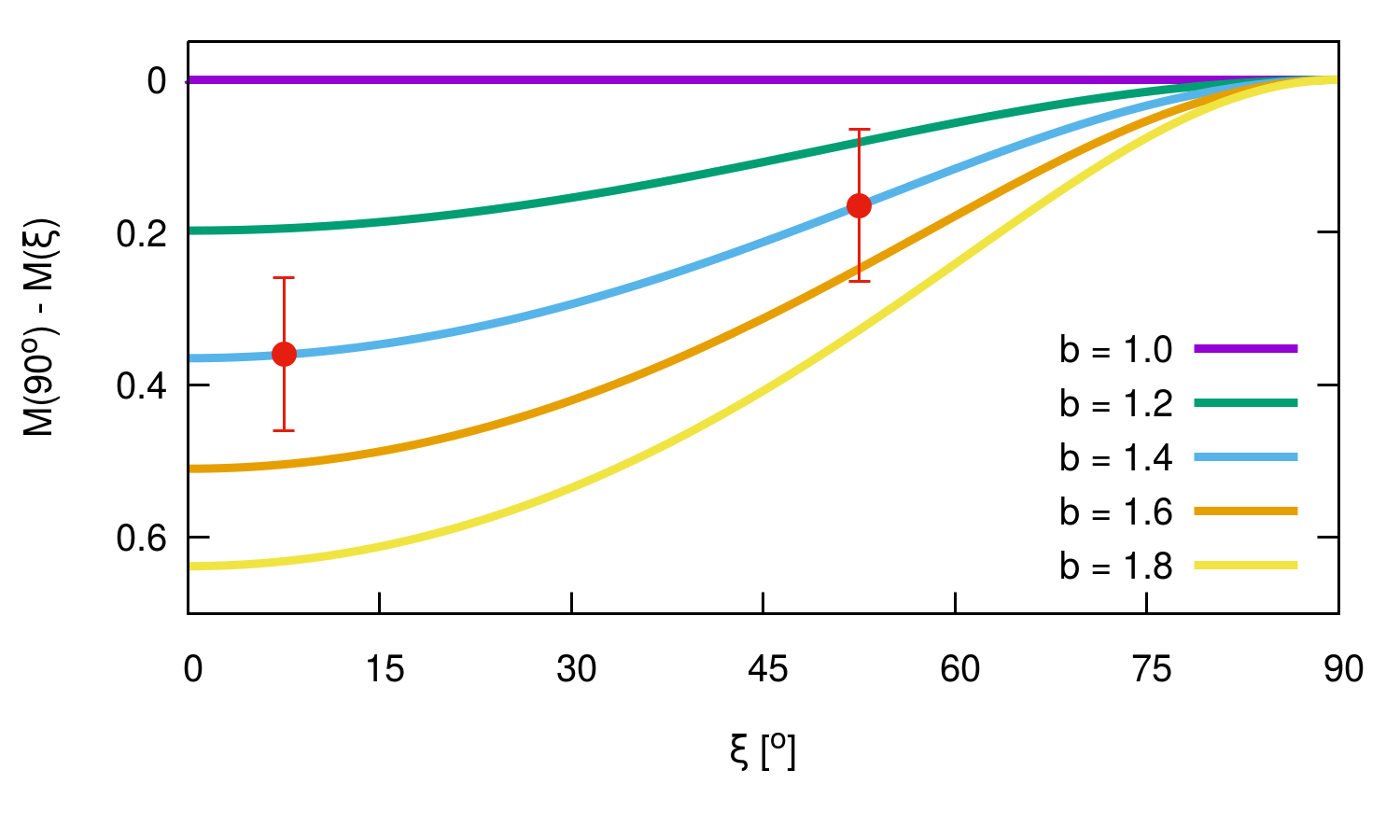}
	\caption{
		The plot of difference in magnitudes (Eq.~\ref{eq:DM}) versus aspect
		angle $\xi$ for different values of semi-axis $b$. The semi axis $c=1$
		by definition. The red dots show
		difference in magnitudes from photometric measurements with magnitude
		precision $\delta m = 0.1$, a tipical precision from
		 surveys \citep{Hanus2011}. The expected precision of upcoming Gaia measurements of
		 $0.001$ mag \citep{Mignard2007} is about the size of the points.
	}
	\label{fig:bees}
\end{figure}

\begin{equation}
	2.5\log d \pm \delta m = 2.5\log d',
\end{equation}
and solving it for $b'$ we get
\begin{equation}
	b'^2_\pm = b^2 \cdot 10^{\pm 0.8 \delta m} + ( 10^{\pm 0.8 \delta m}
	-1)\tan^2(\xi).
\end{equation}
Now, we are able to calculate the relative uncertainty of volume
\begin{equation}
	U(V)/V = \frac{b'_+ + b'_-}{b}
\end{equation}

\begin{figure}
	\includegraphics[width=8cm]{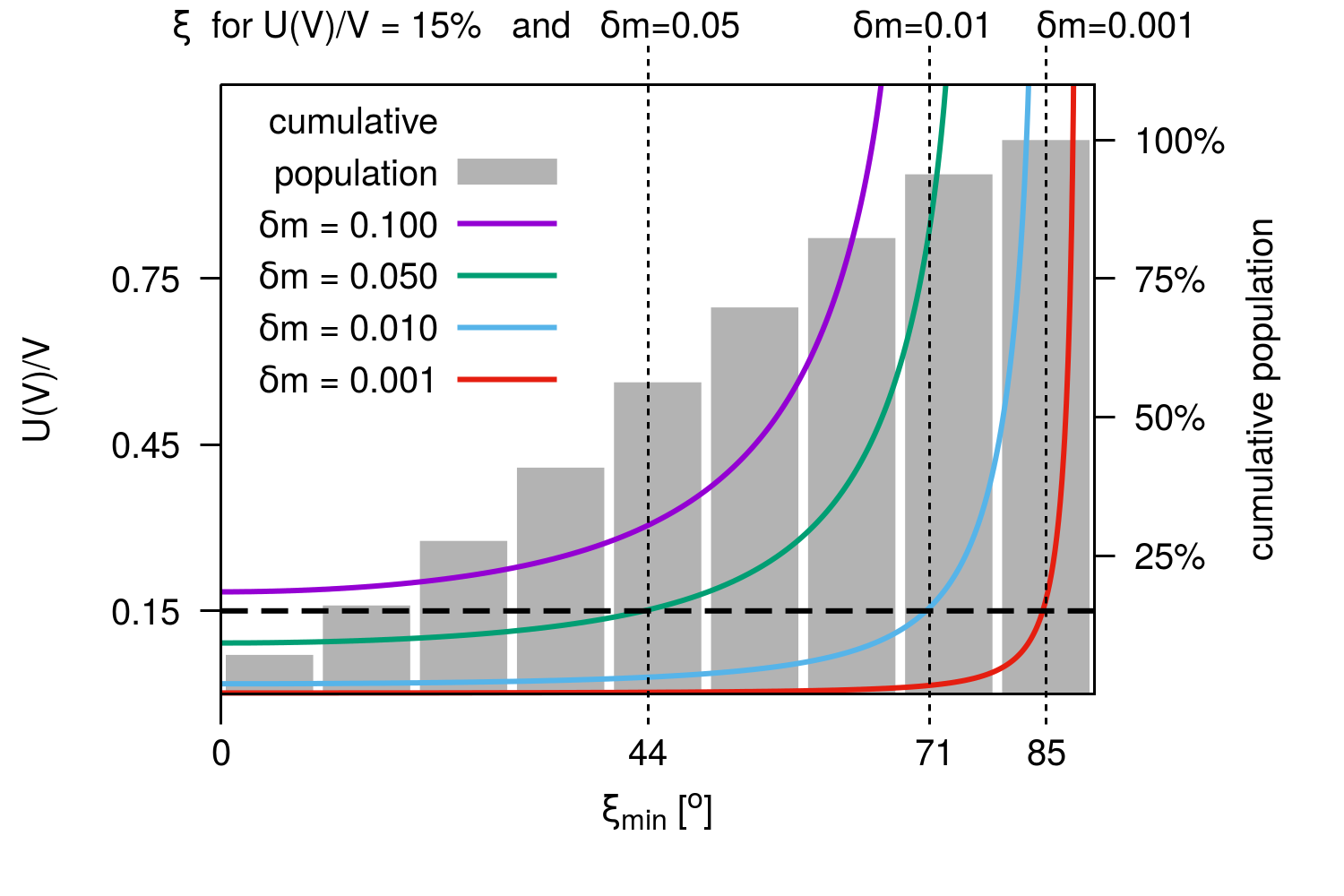}
	\caption{
		Relative volume uncertainty $U(V)/V$ for minimal observable aspect
		$\xi_{\text{min}}$ (colour lines) and the cumulative population of
		asteroids  (gray bars) for which this is achievable. The black dashed
		horizontal line shows 15\% level of volume uncertainty. See text for
		detailed description.
		 }
	\label{fig:chances}
\end{figure}

Fig.~\ref{fig:chances} shows the relative volume uncertainty $U(V)/V$ for
$b=1.2$ and $\delta m = 0.1$, $0.05$, $0.01$ and $0.001$ (colour lines) and the
cumulative fraction of asteroids (gray bars) for which minimal observable aspect
angle $\xi_{\text{min}}$ is achievable. Aspect angles were normalized to $[0\dd,
90\dd]$ range because of the symmetry of ellipsoidal model. The distribution of
spin axis orientations was taken from Asteroid Spin Vectors database
\citep{asv}. It has to be stressed that the population of models and estimates
used to derive those spin properties is biased \citep{Marciniak2015} and so are
the results of our evaluation. Together with the information on the orbits'
inclinations the minimal achievable aspects were calculated neglecting the
longitude of spin axis orientations $\lambda_{pole}$ and the orbits longitudes
of ascending node $\Omega$.

As pointed out by \cite{Carry12b}, when mass of an asteroid is known
precisely (i.e. relative precision better than 20\%) the size estimate is the
limiting factor on the density assessment; the models' volume uncertainties
below 10-15\% are needed to exploit any mass determination. Assuming said volume
accuracy, we can calculate minimal required aspect $\xi_{\text{min}}$ (dashed
vertical lines in Fig.~\ref{fig:chances}) below which it is feasible. It
further allows us to estimate the fraction of asteroid population for which this
is doable.  For magnitude precision $\delta m = 0.1$ this fraction is 0.  When
we increase the accuracy to $\delta m = 0.05$ the fraction is enlarged to 48\%
of the population with $44\dd$  minimal aspect angle required, while with
$\delta m = 0.01$ it is 88\% at $71\dd$ aspect angle.
For $\delta m =  0.001$ volume uncertainty below 15\% can be achieved for almost
all targets. Such accuracy will be feasible for a fraction of asteroid population
using the Gaia observations \citep{Mignard2007}. The outcome, however, depends
on the plethora of factors and is therefore target-specific \citep{Spoto2018}.

It is worth remembering that this estimate is rough, but very informative, as it
shows the restrictions coming from observing geometries, i.e. what is the best we
can hope for. Absolute observations are usually sparse-in-time. Imprecise
rotational period estimates leading to big uncertainties in rotational phase for
a given time, and low precision in magnitude can pose problems as well when
comparing the model to the data. The need for good quality observations is
obvious and we all look forward to getting the precise measurements, like the ones Gaia
mission will provide. Other techniques offering absolute measurements of the
size of a body, i.e. stellar occultation chords and adaptive optics images,
could also be used to properly scale the model in z-axis.  However, the geometry
of such measurements needs to be favorable as well, meaning the aspect of an
asteroid needs to be close to $90\dd$. The number of observations utilizing such
techniques is still very low (an order of magnitude lower compared to lightcurves).

\section{Uncertainty Assessment Method}
\label{sec:method}

The main goal of the method presented in this section is to convert a
deterministic asteroid model into stochastic one through sensitivity analysis.
This modelling-technique independent method is based on creating clones of a
nominal model, introducing changes to its parameters and accepting or rejecting them
based on how they fit the observations. Uncertainties of model parameters' are
derived from per parameter ranges of values found in accepted clone
population. The population of clones inside the confidence level can
subsequently be used to create predictions of observations with probability
associated with simulated data point rather than single value. Only
disk-integrated photometry is considered in this work. The scattering law used
to create synthetic lightcurves is the same as the one used during the
modelling.

\hilight{Although rotational period, phase of rotation for reference epoch and
scattering law are explicitly defined}, the shape parameters describing the same
surface can be defined in many ways, e.g. voxels, spherical harmonics, etc. Each
representation can be translated to a surface composed of triangular facets
defined on a mesh of vertices in 3D space and this scheme is used throughout
this work.

\subsection{Remesh}

\begin{figure}
	\center
	\includegraphics[width=4cm]{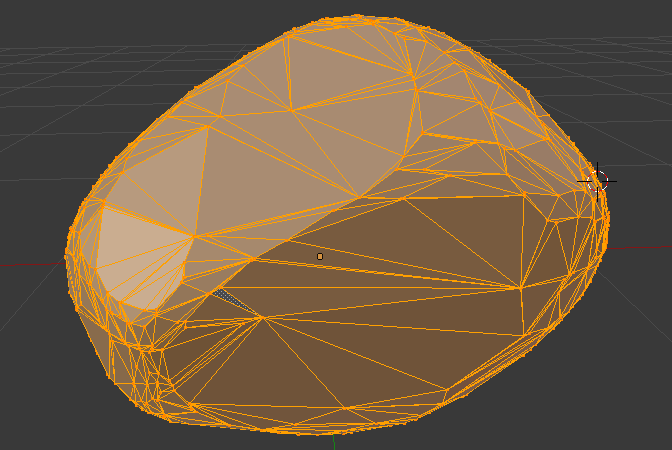}
	\includegraphics[width=4cm]{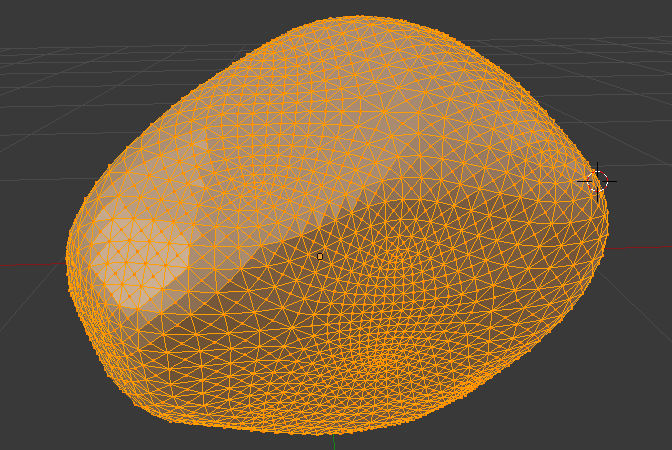}
	\caption{Example shape before (left) and after (right) remeshing operation.
	See text for detailed description.}
	\label{fig:remesh}
\end{figure}

There are many ways to represent and change 3D shapes and surfaces.  Meshes of
vertices and triangular facets defined on them are primarily used in computer
graphics and also in asteroid shape modelling.  When dealing with models from
different asteroid modelling methods the need for unifying the shape
representation arises. Making small, uniform changes in models' parameters is
the key component of the method described in this work, and to assure uniformity
in model's shape changes (guaranteeing good statistics), vertices need to be
evenly distributed. That also requires consistent triangle sizes, lack of
spike-like or huge ones. After remeshing resulting shape stays the same (and
gives the same lightcurves, which is of our main concern here), but is much
easier to work with (Fig.~\ref{fig:remesh}).

The new mesh is achieved by calculating intersection points between model's
surface and equally distributed set of rays originating in the model's centre.
The nature of surface fluctuation creation and their application to the model
(see Sec.~\ref{sec:fluctuations}) require even mesh, no matter the
shape. To get satisfactory level of triangle shapes and sizes uniformity
sufficiently big number of rays is used. To construct the set of rays we start
with the largest platonic solid consisting of 12 vertices and then use surface
subdivision algorithm \citep{Catmull-Clark} to get bigger and uniform meshes.
After 2 iterations we get 242 vertices, and 3842 after 2 more. Triangular
facets are well defined on such mesh which is the obvious benefit of
transforming vertex positions into rays.  Summing up, after remesh the shape
changes are made on the mesh of 3842 vertices with 7680 triangular
facets.

This simple scheme is sufficient for most asteroid shapes although some
modifications would be necessary for strongly non-convex shapes.
Model's reference frame is left intact, i.e. we do not
recalculate it's centre nor apply any rotations.  A model being tested after
remesh operation is called \textit{reference} or \textit{nominal} model
throughout this work.

\subsection{Confidence level of the nominal model}
\label{sec:confidence_level}

Before cloning of the nominal model can begin a confidence level of the nominal
model needs to be established. Later, based on it, the decision whether to
accept or reject a clone is made. For a set of observations, against which the
model is tested, weighted root-mean-square deviation is
calculated:
\begin{equation}
	\label{eq:rmsd}
	RMSD_{ref} = \sqrt{ \frac{\sum_i^N w_i (O_i-C_i)^2}{N \sum w_i } }
\end{equation}
where $O_i$ and $C_i$ denote observed and corresponding synthetic photometric
points and $N$ denotes total number of them. \hilight{The weight $w_i$ is
connected with magnitudes precision of a photometric point $\sigma_i$, i.e. $w_i
= 1/\sigma_i^2$.  }

Next, a standard error of $RMSD_{ref}$ distribution of a form

\begin{equation}
	\mathcal{E} = \frac{RMSD_{ref}}{\sqrt{N-n}}
\end{equation}
with model's $n$ degrees of freedom is used to establish a test for the clones.
Accepted ones need to satisfy the following equation:

\begin{equation}
	RMSD_{clone} \leqslant RMSD_{ref} + \mathcal{E}.
	\label{eq:clone_test}
\end{equation}

Altogether, we are interested in the clones with the level of goodness of the
fit equal or better than the nominal model, as established by $\mathcal{E}$.

Lightcurve comparison is straightforward. Synthetic lightcurves are created for
every corresponding available observation. As the absolute magnitude information
is absent, the synthetic lightcurves are shifted so that average magnitudes of
real and synthetic observations match in each lightcurve.  Then, $RMSD$ value is
calculated per lightcurve according to eq.~\ref{eq:rmsd}.

The absolute sparse photometry is processed differently. The increased
brightness of asteroids evident in observations taken near opposition (i.e.
phase angle less than $8\dd$) depends on spectral type, is non-linear and
strongly depends on on the surface properties \citep{Belskaya2000,
Muinonen2010}. In the context of this work non-linear part of a phase curve does
not contain any crucial, additional information compared to the linear part of a
phase curve. To minimise the risk of bad fitting, and therefore influencing the
results, we omit the data with phase angles smaller than $8\dd$.

For the remaining linear part the $Ax + B$ function is fitted to the data
(separately for real and synthetic observations). The slope parameter $A$ is
calculated simultaneously for all data points, while $B$ is calculated
separately for each of group of data points with similar aspect angles, with
$5\dd$ resolution of the bins. The $RMSD$ value comes from the discrepancy of
$B$ parameters. That way we can search for proper z-scale.

$RMSD_{ref}$ value in eq.~\ref{eq:clone_test} is the sum of separately computed
values for each observation type.

\subsection{Surface fluctuations}
\label{sec:fluctuations}

Introduction of surface fluctuation to the nominal model is achieved by moving
its vertices according to one of precomputed convolution masks. A set of
convolution masks created beforehand enables testing of the uniformity of
changes applied later to any model and is computationally more efficient than
doing it ad hoc. Below we describe how convolution masks are created.

Rather than changing vertex positions independently the hills and concavities of
various shapes and sizes are introduced. First, a vector $v_i$ is randomly
selected and it's position from the centre is randomly changed within $0.5$ to
$1.5$ range of initial distance from body centre. Then, all other vertices are
being altered as well in order to ensure smooth surface. If $\alpha_{ij}$ is an
angle between the two vertices $v_i$ and $v_j$, then the formula
\begin{equation}
	v'_j = v_j A \sin(\alpha_{max}/2) \exp(-2k\alpha_{ij}/\alpha_{max}),
\end{equation}
where $A\in [-0.25,0.7]$, $\alpha_{max} \in [50\degree,90\degree]$ and $k \in
\{1,2,3\}$ are random numbers, will generate hills or concavities of different
width and flatness when applied to vertex distances from the model's centre. The
amplitude $A \sin(\alpha_{max}/2)$ controls the proportion of the base to the
hight of a hill and its range is specifically chosen to eliminate undesired
spikes or deep wells in resulting body.

Changes to a vertex positions in set of convolution masks are generated randomly
but should have uniform distribution not to introduce biases in clones. Because
vertices are not independent of each other, special care has to be taken to
achieve that. In the set of convolution masks certain amplitude of change for
given vertex must appear fixed number of times. With this criterion, when
analysing the probability of a vertex having certain amplitude of change the
same distribution for each vertex is accomplished. The variances of probability
density distributions for vertices' changes differ by only $0.1\%$ -- a
satisfactory level of homogeneity.

\subsection{Procedure}

The algorithm of volume uncertainty assessment described here is based upon the
idea of changing the nominal model's parameters and testing modified models,
i.e. clones, against observations. There are three components considered when
modifying the nominal model: the spin axis orientation, the scale along the spin
axis and surface fluctuations. Combining the scanning of pole position and
$z$-axis scale for each surface fluctuation mask in straight forward manner would
be computationally heavy so a special strategy needs to be applied in order to
assess uncertainty in reasonable time, namely, the procedure consists of two
phases.

In the first stage the nominal model is modified only by applying the z-scale
drawn from $[0.5, 1.5]$ range and by changing the spin axis orientation within
$\pm30\dd$ in both $\lambda$ and $\beta$. For all of the observations synthetic
ones are created and compared.
The criterion used here differs slightly from eq.\ref{eq:clone_test}, i.e.
\begin{equation}
	RMSD_{clone} \leqslant RMSD_{ref} + 3\mathcal{E}.
	\label{eq:clone_test2}
\end{equation}
The surface fluctuations do not alter the volume, lightcurve amplitude and
absolute photometry as much as changing spin axis and z-scale do. This fact is
used to preselect the space of $\lambda$, $\beta$ and z-scale for the second
stage therefore significantly lowering the amount of computation.  Altogether,
this stage generates a list of $\lambda$, $\beta$, z-scale  triples that passed
the criterion.

In the second stage all components are joined together. To produce a clone one
of the surface fluctuation convolution masks is applied to the nominal model.
Subsequently, a triple is randomly drawn from the list created in the previous
stage so that the model can be scaled by z-scale and assigned new spin axis
orientation.  In total there are $1.3\times 10^6$ clones created and each of
them is tested against eq.~\ref{eq:clone_test} thus producing a final population
of accepted clones.

To assess volume and parameters' uncertainties the population of accepted clones
is analysed and extrema are searched for. Reported uncertainties are simply a
range of values in this population for each parameter.  Because of the fact that
the hills and concavities have different influence on the lightcurves the
uncertainties above and below nominal value are reported separately.

\section{Uncertainty propagation}

\subsection{Volume}

Asteroid shape models from lightcurve inversion are dimensionless.
The vertex positions are expressed in the units of the longest vector length
$R_{max}$, and the volume in $R_{max}^3$. By convenience $R_{max}$ is set to 1.
The volume uncertainty for unscaled model -- which comes from
extreme values of accepted clones -- is given as percentage of nominal model's
volume.

Even though the models are unitless the information about the volume can be
extracted by scaling the clones to match the amount of reflected light by the
nominal model which is directly connected to the surface area. Let us consider
3-axial ellipsoid and one photometric point taken at opposition when axis $a$ is
pointing towards, and $b$ and $c$ axes perpendicular to, the observer. When all
three axes are enlarged uniformly in a clone and when it is uniformly scaled so
that the luminosities match we get the same body size and volume. The shape of
the lightcurve from larger clone is exactly the same and there is no reason to
consider this clone to be different from the nominal model. In contrast, if axis
$a$ alone is enlarged, the amount of light stays the same if one uses geometric
scattering law, or changes slightly with more sophisticated one. Therefore,
there is no imperative to scale the whole model (or is to scale it only a little
bit), and so its volume is larger when compared to the reference model. The
opposite would happen if we fix axis $a$ and enlarge axes $b$ and $c$.

This simple example gets very complicated quickly if we consider lightcurves
covering partial or full rotation rather then single points, additional
observations from different positions and phase angles, various spin axis
orientations, etc. We can learn about the volume range from clones statistics
(an average of all the effects), i.e. the scope of solutions allowed by
photometric observations.

\subsection{Scale}

Scaling methods (e.g. fit to stellar occultation chords, thermophysical
modelling, adaptive optics images) provide the scale $S$ by which every vertex
of the model can be multiplied to create a model with dimensions
expressed in physical units. Scale, in that sense, can be understood as
$R_{max}$.

The volume of a scaled model can be expressed as
\begin{equation}
	V_m = V' S^3 \pm u(V_m),
\end{equation}
where $V'$ is a dimensionless volume of unscaled model and $u(V_m)$ is volume
uncertainty.

When scaling the model using different data type from the one used when creating
the model, and such is the case for majority of lightcurve based models, the
resultant scale is biased by the model itself.  This bias can, and should, be
included in the uncertainty of the scale (and scaled volume as well) by
accounting for both measurement and model uncertainties during the scaling
procedure.

When fitting to the occultation cords or adaptive optics images the positions of
the edges of the model on the Earth's surface or plane-of-sky projections can be
transformed from deterministic values into stochastic ones utilizing population
of accepted clones created during uncertainty assessment procedure.

In case of observations in thermal infrared, which are not so sensitive to the
details of the shape \citep{Hanus2015}, one could use the information about
allowed range of the z-scale as well as spin axis orientation uncertainty when
creating thermophysical models. Granted that observations in thermal infrared
provide absolute fluxes they are sensible to the extent of the body along the
spin axis. Changes in z-scale and spin axis will be apparent in flux -- aspect
angle relation and could provide better fits to the data when they deviate from
a nominal shape model.

\subsection{Rotational period and phase of rotation}

Rotational period is not a free parameter of a clone.  The rotational phase for
reference epoch  uncertainty $u(\gamma_0)$ (i.e.  uncertainty of the rotational
phase at $JD_0$) depends on rotational period uncertainty, data points precision
and density of the lightcurves and observations' time span. For each clone the
period $P$ and $\gamma_0$ are calculated separately from the best fit to
synthetic lightcurves.  Discrepancy of the values in the accepted clones
population for both are treated as their uncertainties.  The nature of
$\gamma_0$ uncertainty is different from the rotational phase uncertainty mainly
because it does not accumulate with time.  These values have to be added
together in order to facilitate total rotational phase uncertainty.

\begin{figure}
	\begin{center}
	\includegraphics[width=8.5cm]{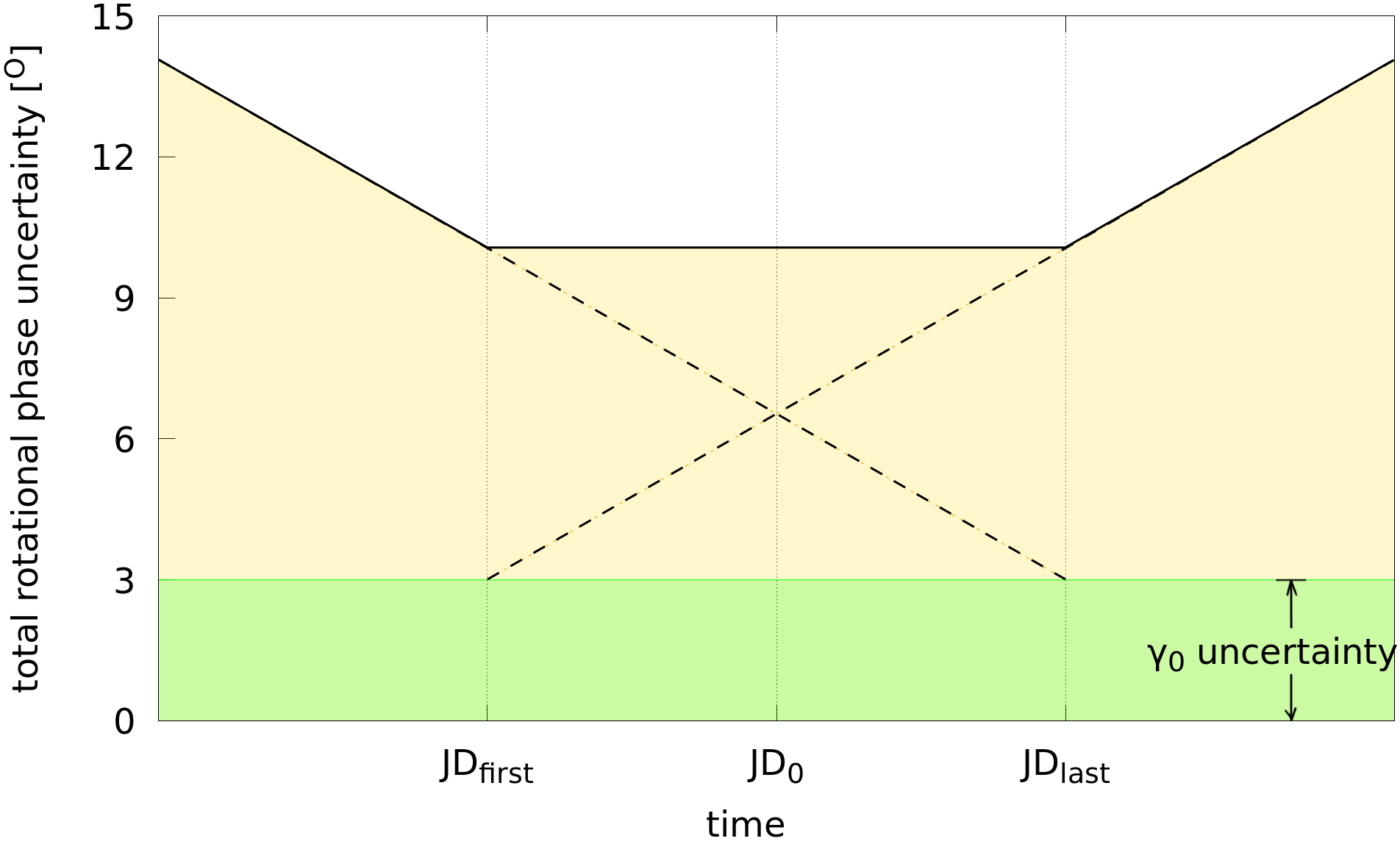}
	\end{center}
	\caption{The plot of total rotation phase uncertainty. In this example
		$\Delta T_{obs} = 38.5$ years,
		the uncertainty of rotation
		phase for reference epoch $JD_0$ is $u(\gamma_0) = 3\dd$
		and the period uncertainty $u(P) = 1.5\times 10^{-6}~h$ .}
	\label{fig:period_uncertainty}
\end{figure}

The uncertainty of rotational period propagates into phase of rotation.  The
rotational phase uncertainty should stay on the same level inside the interval
between the first and the last observation because the period was established
using all the available lightcurves at once. Outside this time interval
rotational phase uncertainty increases linearly with time from the last
observation. To facilitate the above in one formula we put the reference epoch
right in the middle of observing time span, i.e. $JD_0 = (JD_{last} +
JD_{first})/2$, and then calculate

\begin{equation}
	\begin{aligned}
		u(\gamma) &= \frac{2\pi}{P^2}
		\left[a + b + max(-a + b, 0) \right] + u(\gamma_0), \\
		a &=  |u(P) \Delta t|,					\\
		b &=  \frac{1}{2}u(P) \Delta T_{obs}
	\end{aligned}
\end{equation}
where $P$ is a rotational period, $\gamma$ is a rotational phase,
$\gamma_0$ is a rotational phase for reference epoch $JD_0$,
$\Delta T_{obs} = JD_{last} - JD_{first}$ is an observations' time span,
$\Delta t = t - JD_0$ is a distance in time from the reference epoch, and
$u(x)$ denotes uncertainty of $x$. The $max$ function returns larger of the two
arguments. An example plot of this function is shown in
Fig.~\ref{fig:period_uncertainty}.

\section{synthetic example}
\label{sec:synthetic_example}

In order to test the method and show how the dataset influences the model
uncertainty we created 3 sets of synthetic lightcurves of a 3-axial ellipsoid
with $a/b = 1.5$ and $b/c = 1.14$.  These lightcurves served as reference
observations. The body used for testing was the same ellipsoid but with two
craters put at ($40\dd$, $0\dd$) and ($305\dd$, $-40\dd$) longitude and latitude
coordinates. We wanted to test whether the craters -- treated as fictitious
features of the body -- will be detected as the surface areas with large
uncertainty. Moreover, gaps in the rotational phase were introduced in one of
the sets and were expected to influence the uncertainty as well. In summary,
modified ellipsoid with craters was being altered during the process and
compared with lightcurves of an ellipsoid.

\subsection{Lightcurves}
\label{sec:lcs}

3 datasets were created, A, B and C. Each consist of $8$ apparitions evenly
distributed on the orbit (Fig.~\ref{fig:app}). The target's and the observer's
orbits were circular with semi-major axes $a_{\text{target}}=3$ and
$a_{\text{obs}}=1$; apparitions were one year apart. The rotational period of
the body was $P=4.12345$h and pole orientation was $\lambda=0\dd$,
$\beta=45\dd$. For such pole latitude the whole body is visible for the observer
if the target's and observer's orbits are coplanar and the apparitions are
evenly distributed. Had $\beta$ been $\pm 90\dd$ all the lightcurves would have
looked the same and more than one lightcurve would have helped only for
rotational period estimation, whereas $\beta=0\dd$ would produce at least two
flat lightcurves with no shape information in them. There was no Gaussian noise
added to the photometric data points.

\begin{figure}
	\includegraphics[width=8cm]{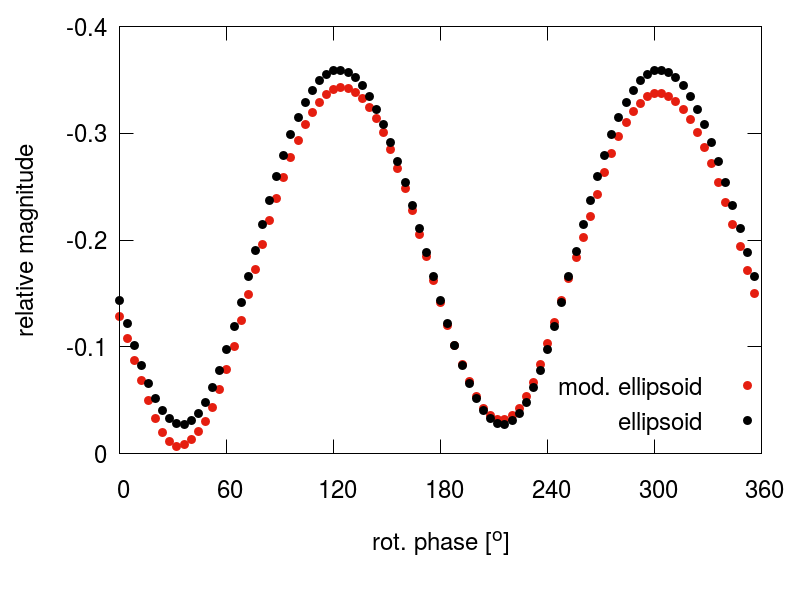}
	\caption{Comparison of synthetic lightcurves of 3-axial ellipsoid and
		ellipsoid with two craters with the largest RMSD value of $0.0155$.
		Lightcurves were obtained at $18.5\dd$ phase angle at $6^{th}$
		apparition.  (see Fig.~\ref{fig:app}) }
	\label{fig:lc_comparision}
\end{figure}

Dataset A consisted of 24 lightcurves covering full rotational period. Each
apparition had 3 lightcurves with phase angles $18.5\dd$ pre opposition, $0\dd$
and $18.5\dd$ post opposition. Dataset B was reduced to 8 lightcurves (one per
apparition) with $18.5\dd$ pre opposition phase angle. Comparison of lightcurves
from dataset A with the greatest RMSD is shown in Fig.~\ref{fig:lc_comparision}.

\begin{figure}
	\includegraphics[width=8cm]{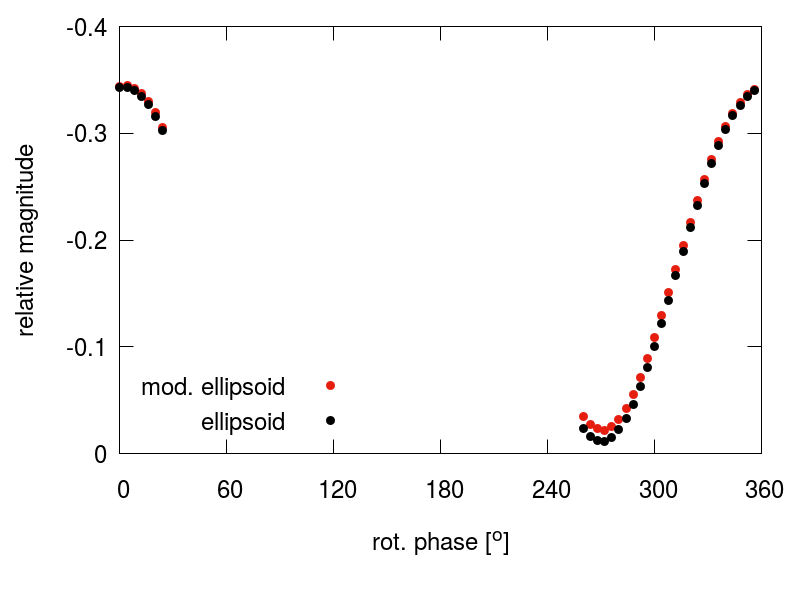}
	\caption{Comparison of synthetic lightcurve of 3-axial ellipsoid and
		ellipsoid with two craters with limited viewing geometries taken at
	$18.5\dd$ phase angle at $8^{th}$ apparition with $\lambda_{obs}=315\dd$
	(see Fig.~\ref{fig:app}).}
	\label{fig:example_lc}
\end{figure}
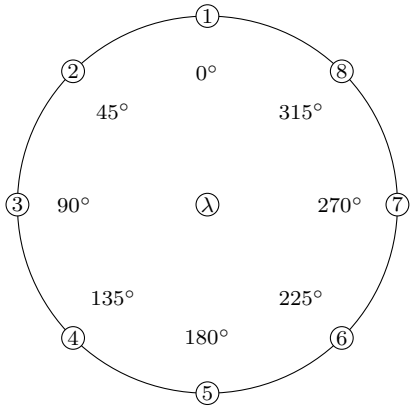
\begin{figure}
    \begin{center}
        \begin{tikzpicture}
            \def\R{2.5}

            \draw (0,0) circle (\R);
            \node(1) at (0,0) {$\Huge{\lambda}$};
            \draw (0,0) circle (0.06*\R);

            \foreach \x in {0,45,..., 315}
            {
                \pgfmathtruncatemacro\y{\x/45 + 1}
                \fill[fill=white] (\x +90:\R) circle (0.06*\R);
                \draw (\x +90:\R) circle (0.06*\R);
                \node(1) at (\x +90:\R) {\y};
                \node(1) at (\x +90:0.7*\R) {$\x^{\circ}$};
            }
        \end{tikzpicture}
        \end{center}
        \caption{
		The image shows the distribution of apparitions (with the observer in
		the centre of the graph) in heliocentric reference frame. See text for
		detailed description.
	}
    \label{fig:app}
\end{figure}

Dataset C was based on dataset B, but the coverage of the rotational phase was
limited, namely, lightcurves covered $3/8$ of the rotational period. One of the
lightcurves is presented in Fig.~\ref{fig:example_lc}. In the body's reference
frame the observer was situated along longitudes from $270\dd$ to $45\dd$.  The
choice of longitudes' range was motivated by the positions of the craters which
were desired to be covered in the lightcurves.

\subsection{Results and discussion}

\begin{figure}
    \begin{center}
	\includegraphics[width=8.5cm]{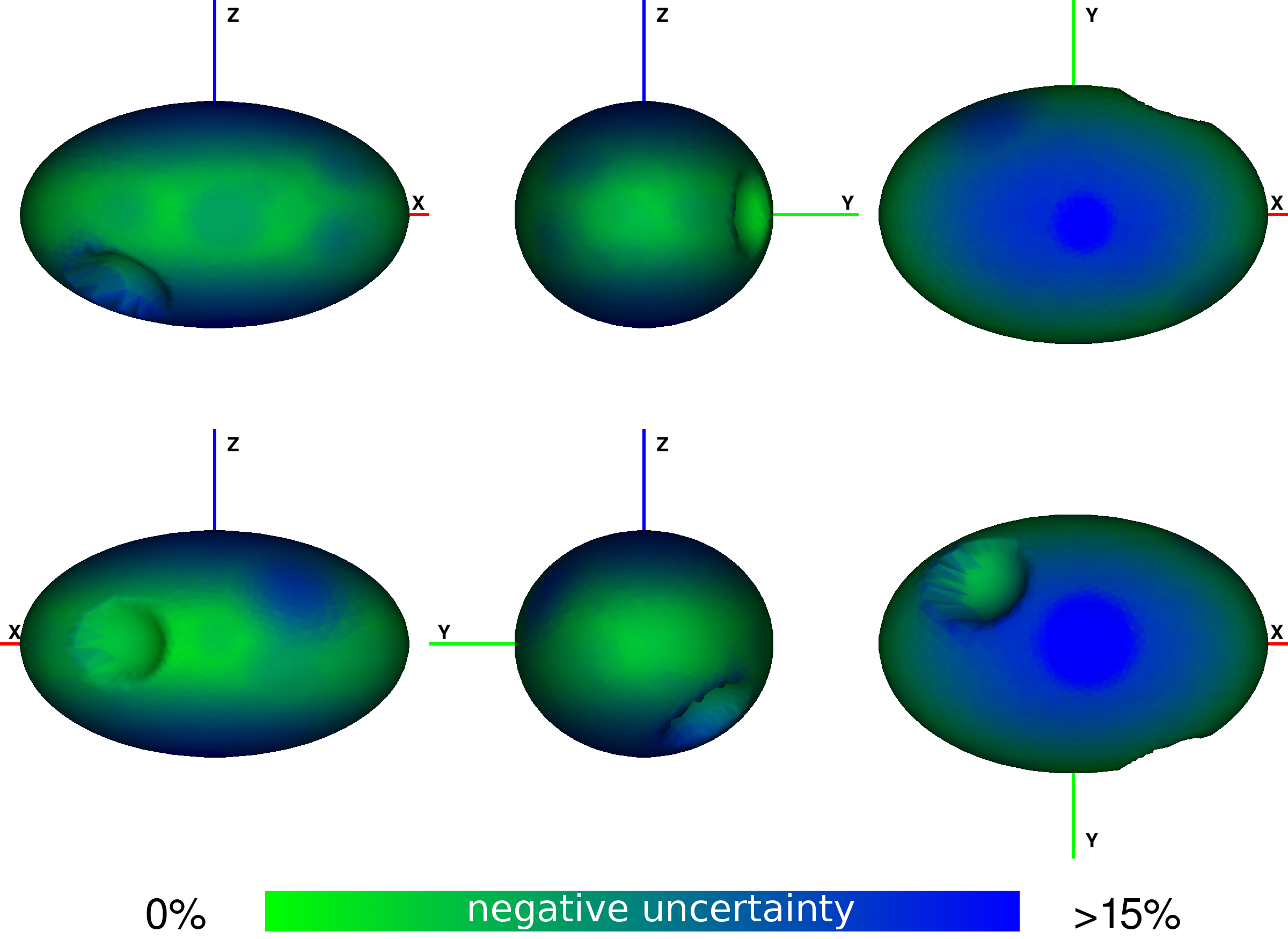}

	\vspace{0.5cm}

	\includegraphics[width=8.5cm]{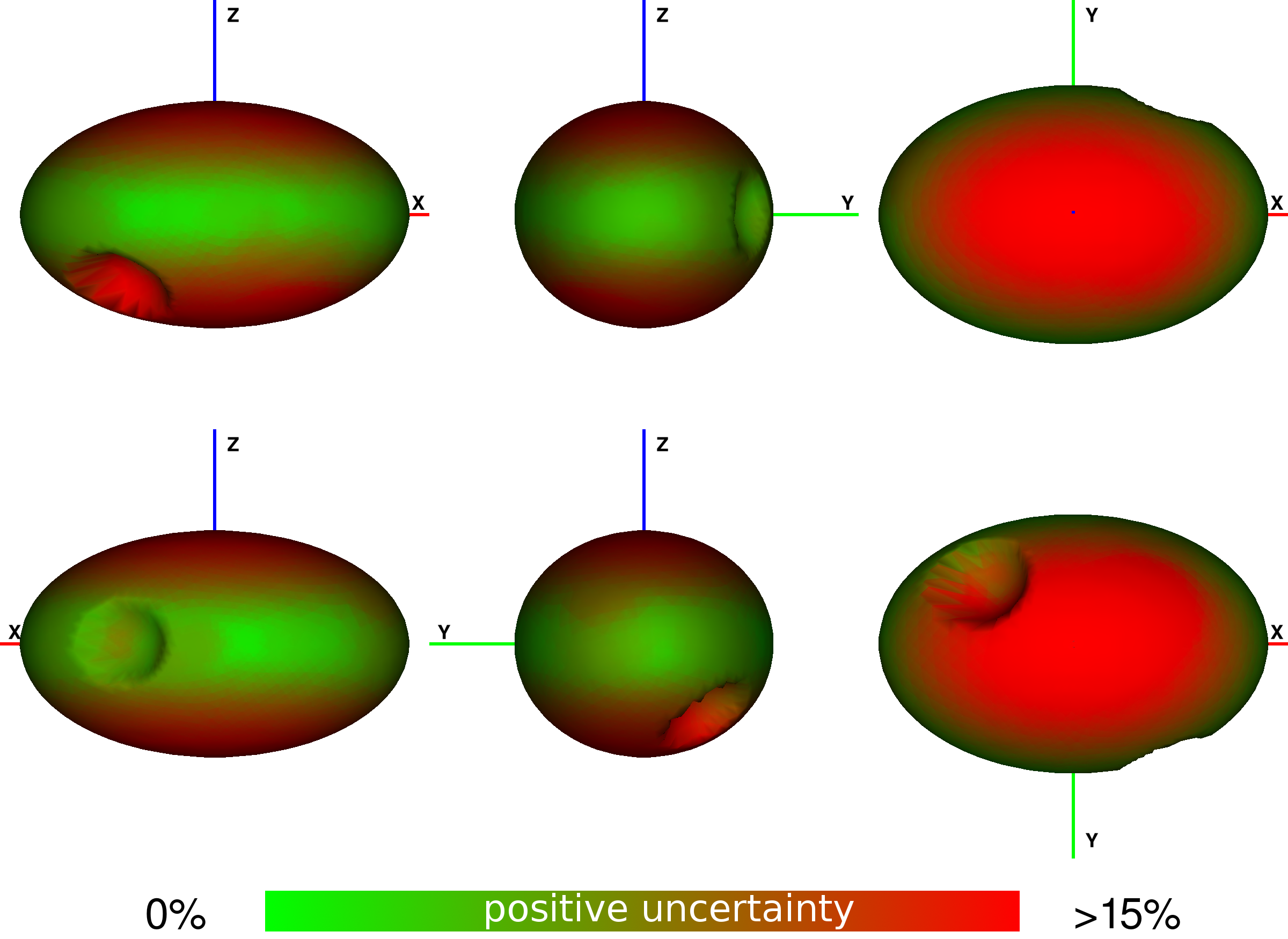}
    \end{center}
	\caption{
		Projections of the target with colour-coded uncertainty information in
		regard to dataset C.
	The uncertainty values come from the range of values found in accepted clones
	population. Top figure shows uncertainty due
to changes that create concavities (blue colour) while the bottom shows the
hills (red). The scale of the values is the same for the two uncertainty maps, the
brightest blue/red showing the values $>= 10\%$ of $R_{max}$.}
	\label{fig:maps}
\end{figure}

\begin{figure}
    \begin{center}
	\includegraphics[width=8.5cm]{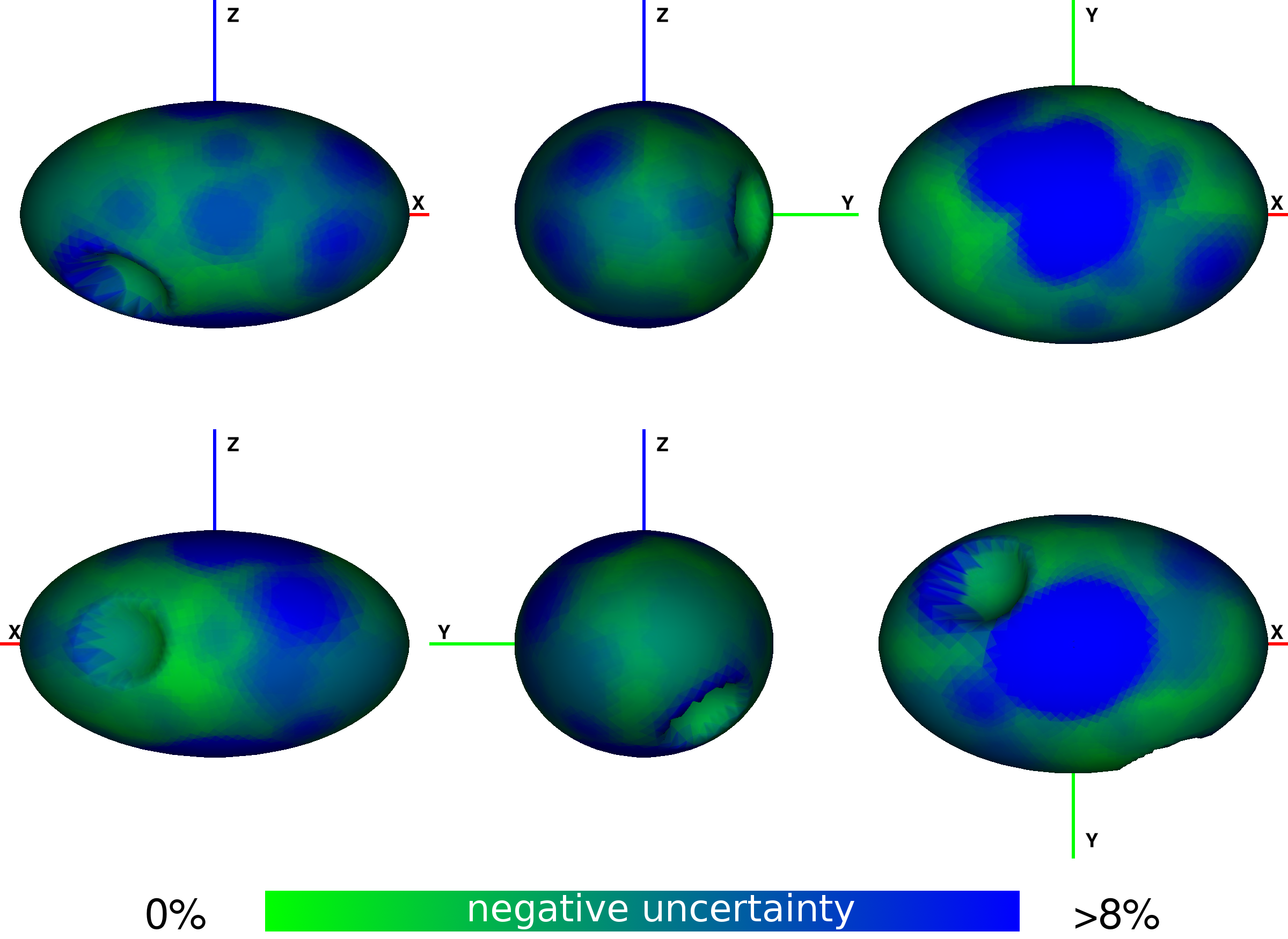}

	\vspace{0.5cm}

	\includegraphics[width=8.5cm]{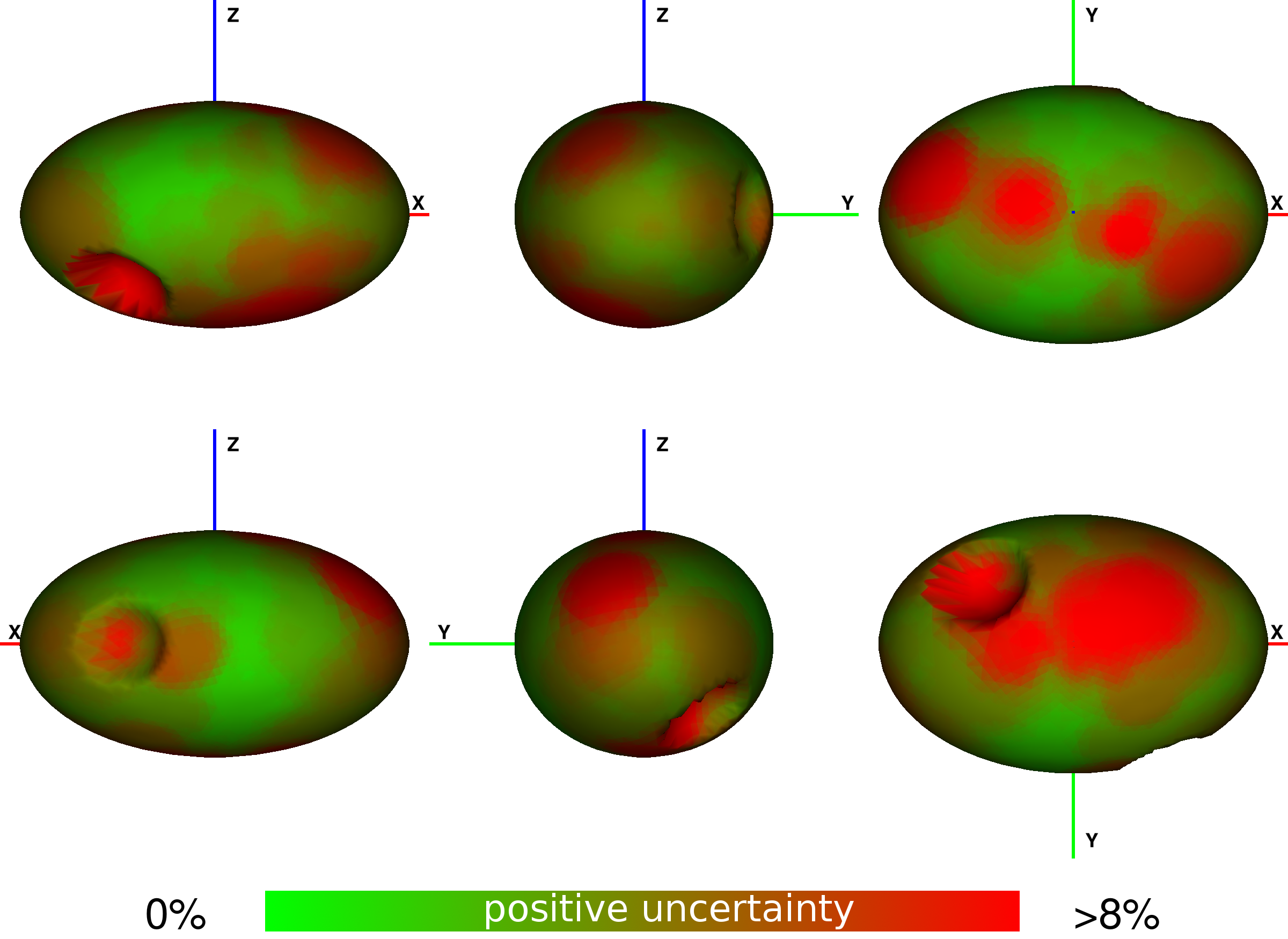}
    \end{center}
	\caption{Projections of the target with colour-coded uncertainty information in
		regard to dataset C and with subtracted z-scale value, thus showing only
		surface fluctuations part of uncertainties. See text and
	Fig.~\ref{fig:maps} for detailed description.  }
	\label{fig:maps_flux}
\end{figure}

To compute uncertainty of the volume we searched for the largest and the
smallest body among the accepted clones. Parameters' uncertainties were acquired
similarly, i.e. by searching for the minimal and maximal values of given
parameter in the population as was the case for the rotational period
uncertainty as well.  For dataset A there were $77162$ clones out of $1.3 \times
10^6$ that satisfied Eq.~\ref{eq:clone_test}, while for datasets B and C there
were $65474$ and $21435$ of them. The results are summarised in
Tab.~\ref{tab:synthetic_results}.

The reason why the number of accepted clones decreased as datasets B and C were
depleted is that the bigger range of spin axis positions were allowed due to
larger standard error $\epsilon$ of nominal model's initial fit.  During initial
scan and using Eq.~\ref{eq:clone_test2} the range of spin axis orientation and
z-scales is preselected is wider as $\epsilon$ gets bigger.  Because the number
of clones in kept constant, after introducing surface fluctuations more clones
were subsequently rejected by the criterion given by Eq.~\ref{eq:clone_test}.
The bigger the $\epsilon$ is, the worse -- or more quantized -- statistics we
get. The number of accepted clones is however still big enough, that parameters'
and volume's uncertainties are not affected much when minimal and maximal values
from accepted clones population are used to determine uncertainties. However,
the chances to miss an important outlier increase.

In all examples the biggest uncertainties of shape parameters were correlated
with the z-scale. As shown in Fig.~\ref{fig:maps} the areas near the poles
dominate at the level of $18\%$ for positive, and $21\%$ for negative values.
In all datasets the surface fluctuations induced volume changes smaller than
changing the spin axis orientation and z-scale did.  The difference in
volume between the unmodified ellipsoid and the one with craters was $2\%$.
Lightcurves from dataset C yielded, unsurprisingly, the biggest total volume
uncertainty of $52\%$. Full lightcurves in dataset B resulted in lower volume
uncertainty of $20\%$, whereas the richest dataset A resulted in $12\%$.

Further analysis indicates that the uncertainty in the examples can be divided
into two categories: the small effect connected with surface fluctuations and
the large effect connected with the z-scale alone.  Shown in
Fig.~\ref{fig:maps_flux} are projections of the test body neglecting the
z-scale. To achieve that, parameters were referenced not to the nominal body,
but to the body already scaled in z-axis before surface fluctuations were added.
For negative part, largest values were still situated near the poles and dropped
to $18\%$ for negative part. The viewing geometries allowed those areas to be
pushed inside without affecting visible cross-section and thus the lightcurves.
As for the positive uncertainties, the maximal values of $15\%$ were situated at
the positions of the craters. The surface inside the craters was allowed to move
outwards without changing the lightcurves meaningfully until it reached the
level of the unmodified ellipsoid surface. Going further changed the silhouette
of the body influencing the lightcurves more dramatically and leading to larger
and unacceptable RMSD values. The uncertainties measured as percentage of
$R_{max}$ along the ray originating in the body'scentre correspond to the
craters' depth measured that way.

The \textit{geometric scattering law} can be treated as a first approximation of
any more sophisticated one. The area of the body's projection on the observer's
field of view has by far the biggest influence on the amount of reflected light.
That means, that elements perpendicular to the observer change the amount of
light the most if moved (especially outwards). Even though observer saw only
$135\dd$ of the rotation phase almost the whole body was represented in the
lightcurves in that sense. The parts that were not have much larger uncertainty
(up to $8\%$, besides the craters). The gaps in observations allowed
\textit{invisible} parts of the surface to fluctuate more which is apparent
along the line from $315\dd$ to $135\dd$ longitudes on the body.  The
spotted-like nature of uncertainty values, rather than uniform patch, is due to
other effects connected with the actual distribution of apparitions on the orbit
and the inclination of the pole. Also, the initial discrepancy between the
modified and unmodified ellipsoids' lightcurves (that define the confidence
level $\epsilon$) allowed for surface changes away from the craters and
unobserved parts.

\begin{table}
	\caption{
		Compilation of results for 3 different datasets. Here we report
		uncertainties of volume $V$, rotational phase for reference
		epoch $\gamma_0$, rotational period $P$ and spin axis coordinates
		$\lambda$ and $\beta$.
			}
\begin{tabular}{ccccccc}
	\hline
	\hline
	\textbf{dataset} & $u(V)$[\%] & $u(\gamma_0)[\dd]$ & $u(P)$[h] & $u(\lambda)[\dd]$ & $u(\beta)[\dd]$ \\ \hline
	A & $^{+5}_{-7}$   & $^{+1}_{-1}$ & $10^{-6}$          & $^{+1}_{-1}$ & $^{+1}_{-2}$ \\	\hline
	B & $^{+5}_{-15}$  & $^{+1}_{-1}$ & $3 \times 10^{-6}$ & $^{+2}_{-1}$ & $^{+1}_{-4}$ \\	\hline
	C & $^{+30}_{-22}$ & $^{+2}_{-2}$ & $2 \times 10^{-6}$ & $^{+3}_{-2}$ & $^{+8}_{-8}$ \\	\hline
\end{tabular}
	\label{tab:synthetic_results}
\end{table}

\section{Uncertainty assessment of selected asteroid models}
\label{sec:real_examples}

The method was applied to models of (21) Lutetia, (89) Julia, (243) Ida, (433)
Eros, and (162173) Ryugu. Except for Julia, which has adaptive optics images
available, each target has been visited by a spacecraft mission resulting in
either flyby images or full 3D models that allow comparison of the models
derived from photometric data together with their uncertainties. \hilight{For
testing we chose available lightcurve-only based models of those targets.} The
results are summarised in Tab.~\ref{tab:results}.

Absolute observations from \cite{Oszkiewicz2011} were used for all of the
objects. Their precision was formally at the level of $0.01$ mag. Additionally,
for Lutetia, Julia and Ida the Gaia Data Release 2 \citep{GaiaMission, GaiaDR2}
were also used. The coverage of aspect angles is shown in
Fig.~\ref{fig:absolute_aspects}.  \hilight{We removed data with phase angles
less than $8\dd$ (see Sec.~\ref{sec:confidence_level}). This resulted in
removing 20\%, 8\%, 38\%, 2\% and 9\% of the data points for Lutetia, Julia,
Ida, Eros and Ryugu, respectively. This did not compromise available geometries
(i.e. aspect angles) as each apparition had large span of phase angles.}

\input{img/tab_results}

\subsection{(21) Lutetia}
\input{img/tab_lutetia_obs}

Asteroid (21) Lutetia has been imaged by Rosetta mission during  10 July 2010
flyby on its way to comet 67P/Churyumov-Gerasimenko. Combining
spectrophotoclinometry technique used on flyby images with inversion of
photometric lightcurves and adaptive optics images, 3D shape model has been
created by \cite{Sierks2011}. Only the northern hemisphere was observed during
the flyby hence the need for external data to complete unseen half of the body.
Later comparison of lightcurve and adaptive optics based model form KOALA method
\citep{Carry2010b, Drummond2010} with flyby model yielded very good agreement of
those models, volume discrepancy (based on equivalent sphere diameters) being
better than 10\% \citep{Carry12}.

As our method considers disk-integrated data only we used lightcurve-only based
convex model of Lutetia \citep{Torppa2003}. The set of observations is shown in
Tab.~\ref{tab:lutetia_lc}. Upon visual inspection the convex shape model is in
very good agreement with the flyby model, especially in $xy$ plane.
Nonetheless, convex nature of the model and it being based on lightcurve data
prevent the model to represent local topographic features.

\begin{figure}
	\includegraphics[width=8cm]{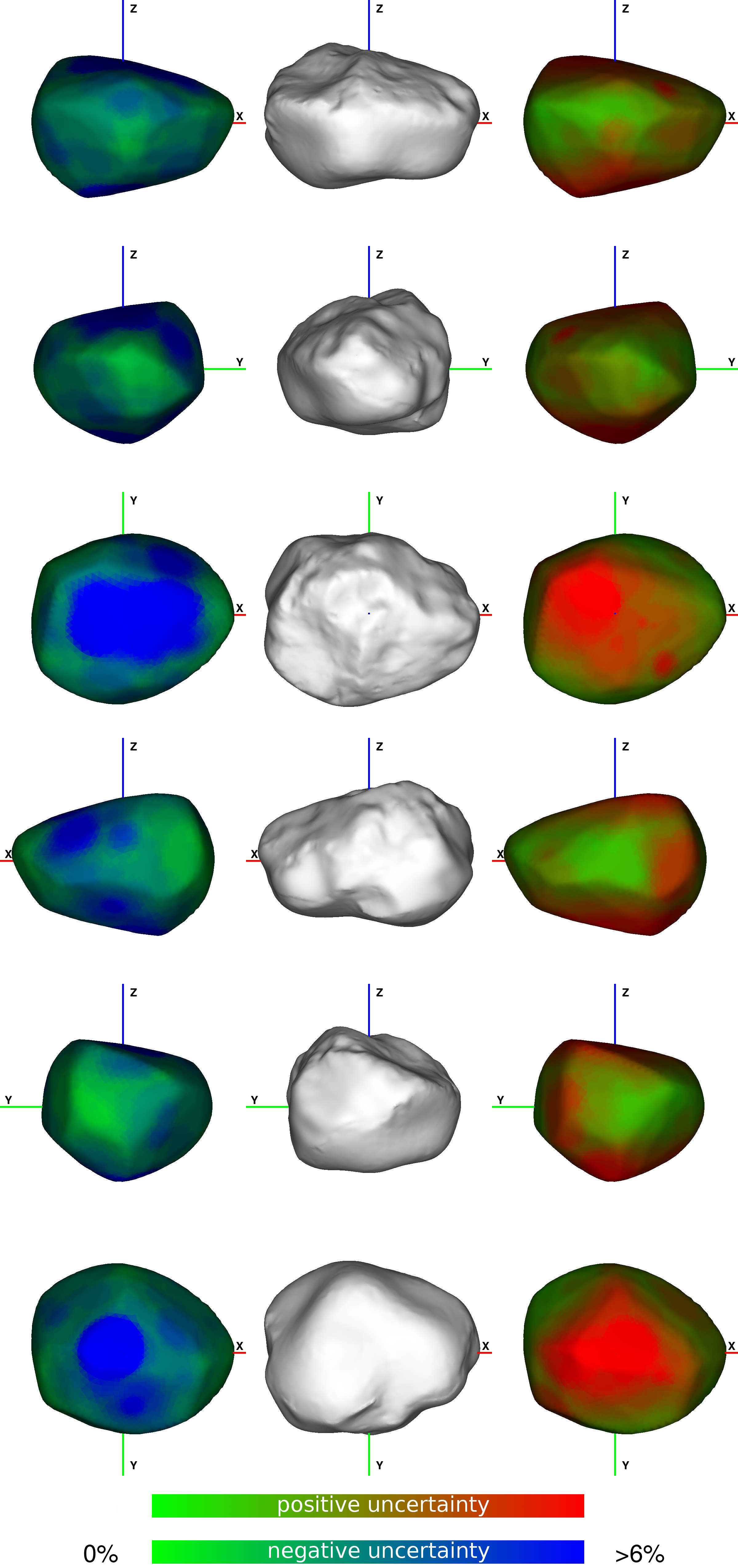}
	\caption{
		(21) Lutetia model \citep{Torppa2003} projections  showing
		uncertainty values  (left and right columns) with comparison to Rosetta
		mission flyby, lightcurve
		and adaptive optics based model (centre column) \citep{Sierks2011}.
	}
	\label{fig:lutetia_projections}
\end{figure}

The analysis yielded maximum parameters' uncertainty of $-13\%$ and $10\%$ of
$R_{max}$ with well constrained spin axis orientation (uncertainty blow $6\dd$).
Convex and flyby model projections comparison is presented in
Fig.~\ref{fig:lutetia_projections}. The biggest values of uncertainties
correspond to purely constrained z-scale in lightcurves. Local uncertainties
correspond nicely with the parts of the body that shows disagreement with the
flyby model. What is more, the parts that do agree (e.g. silhouette in $xy$
plane) are also well constrained by observational data.

\subsection{(89) Julia}
\input{img/tab_julia_obs}

\begin{figure*}
	\includegraphics[width=16cm]{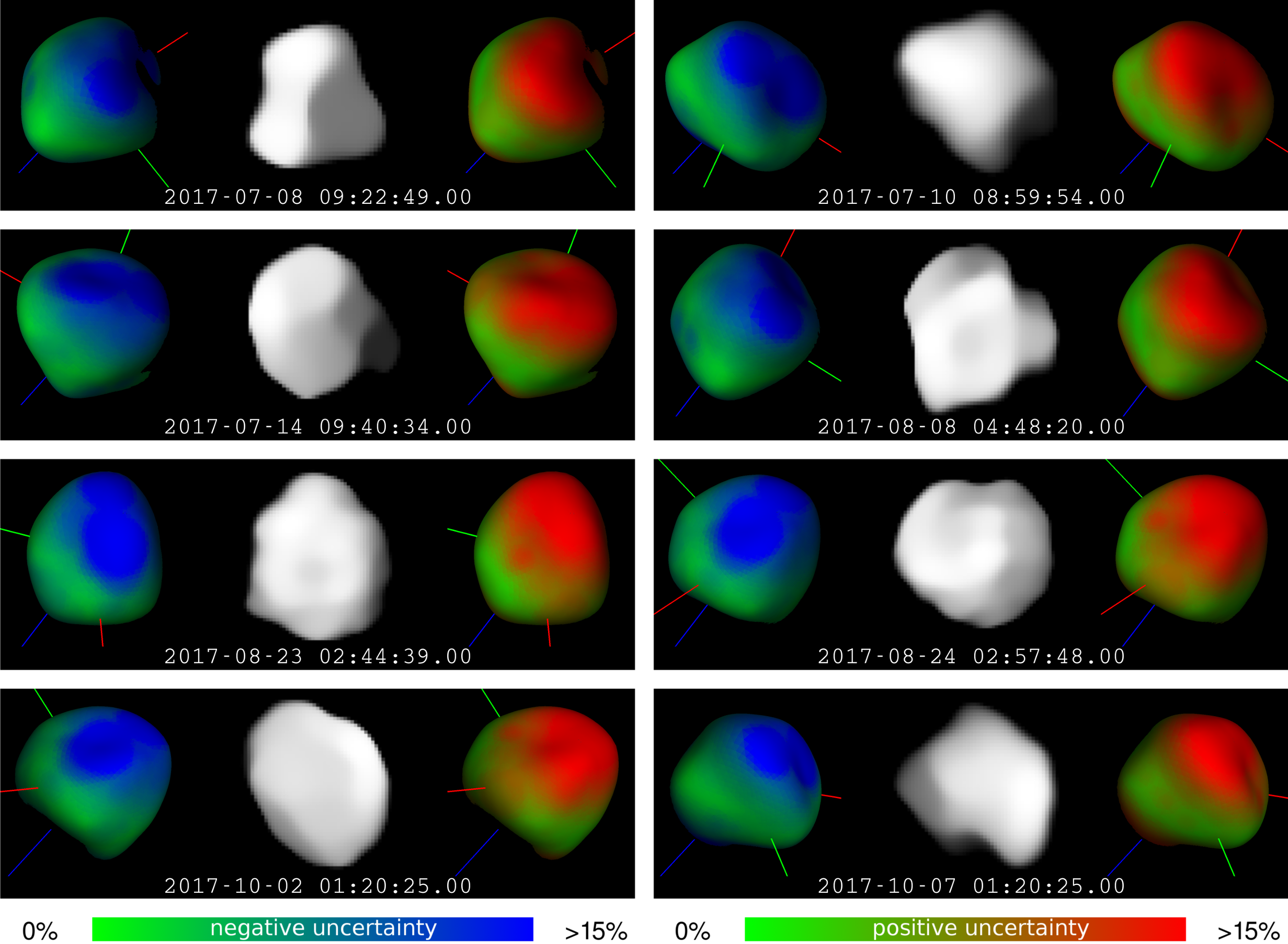}
	\caption{
		Comparison of (89) Julia obtained with SAGE method \citep{SAGE}
		(left and right) with
		adaptive optics VLT/SPHERE images \citep{Vernazza2018} (centre).
		The model axes $x$, $y$ and $z$ are represented by red, green and blue
		lines, respectively.}
	\label{fig:julia_vlt}
\end{figure*}

Although asteroid (89) Julia was not visited by any spacecraft mission it has
available set of adaptive optics images \citep{Vernazza2018} from limited
geometries with aspect angles ranging from $123\dd$ to $140\dd$ and revealing
mostly southern hemisphere of the body. The non-convex
model created for the purpose of this work is based on lightcurve data only
(Tab.~\ref{tab:julia_lc}) and was obtained using SAGE method \citep{SAGE}.
Adaptive optics images were used to compare the model and
uncertainties. The comparison of the model with uncertainties to VLT/SPHERE
images can be seen in Fig.~\ref{fig:julia_vlt}.

\hilight{The non-convex SAGE model did not represent big concavities visible in
the images very well. Nonetheless, the model agrees with disk-resolved images
within  uncertainties (maximal values for parameter were $-20\%$ and $22\%$).
The parts where discrepancies are the biggest had the largest uncertainty
values, which indicates limited information content in the lightcurves.  }

\subsection{(243) Ida}
\input{img/tab_ida_obs}

Asteroid (243) Ida was observed by Galileo spacecraft on 28 August 1993.  The
images obtained during flyby were used to derive 3D shape model
\citep{Thomas1996, Stooke2016}. Convex, lightcurve  based model (see
Tab.~\ref{tab:ida_lc} for description of used observations) of Ida was used
\citep{Hanus2013b}  to assess uncertainties and compare to flyby model.

\begin{figure}
	\includegraphics[width=8cm]{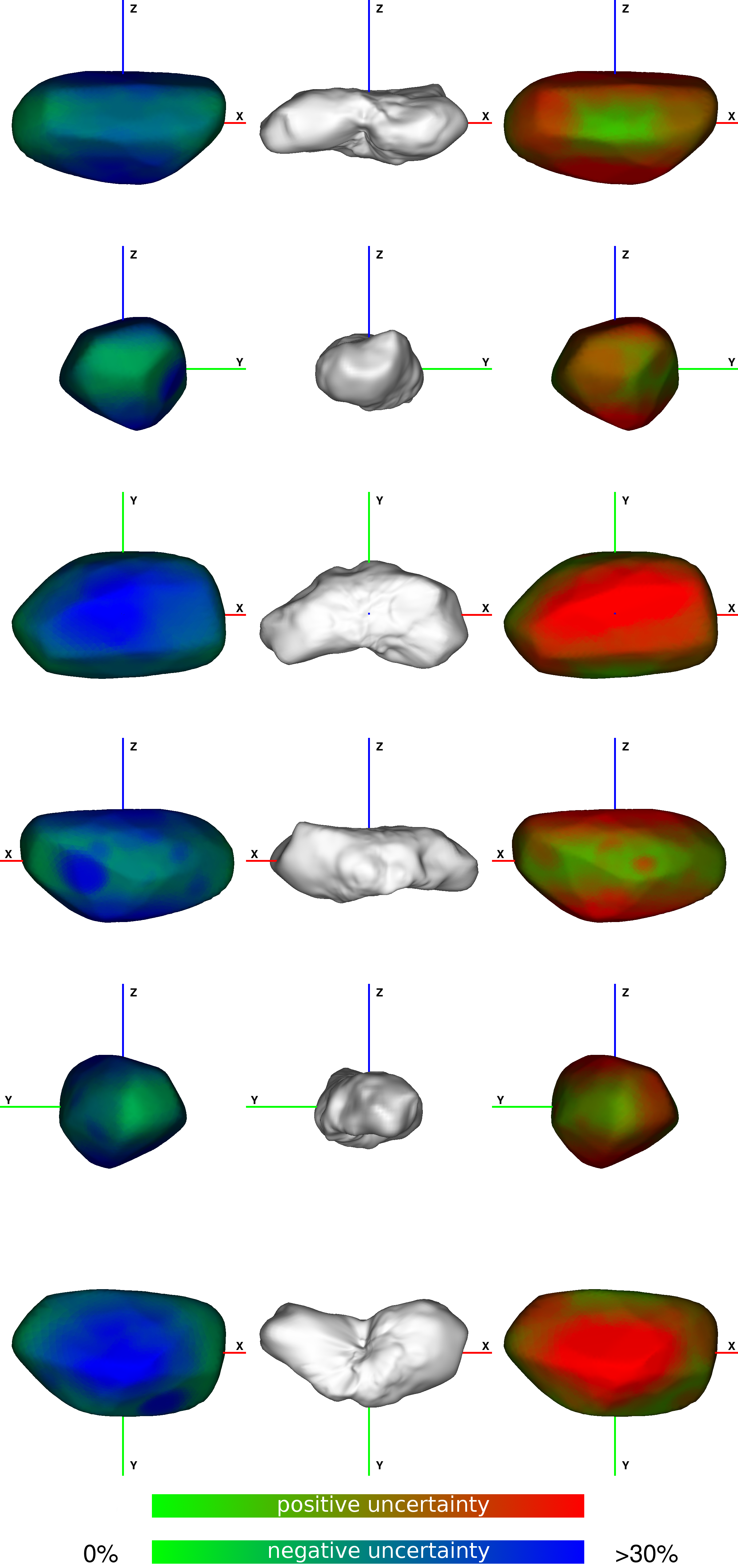}
	\caption{(243) Ida model \citep{Hanus2013b} projections showing uncertainty
	values  (left and right columns) with comparison to Galileo mission flyby
	based model  (centre column) \citep{Stooke2016}. }
	\label{fig:ida_projections}
\end{figure}

Upon first look at images in Fig.~\ref{fig:ida_projections} the convex model is
too extent in z-axis, which is reflected in large ranges of parameter
uncertainties ($-35\%$ for negative and $50\%$ for positive uncertainties) and
in volume uncertainty of more than $100\%$. The model, being convex, does not
reproduce concavities, but negative uncertainties are big in the areas where
flyby model have them. Regions that do match flyby model well have small
uncertainties.

\subsection{(433) Eros}
\input{img/tab_eros_obs}

(433) Eros is an object from Near Earth Asteroid population that has been
extensively studied in the past both remotely and \textit{in situ}. The NEAR
Shoemaker probe, which started orbiting Eros in year 2000 and ended the mission
by landing on the surface in 2001, delivered a detailed 3D shape model
\citep{Zuber2000}.

\begin{figure}
	\includegraphics[width=8cm]{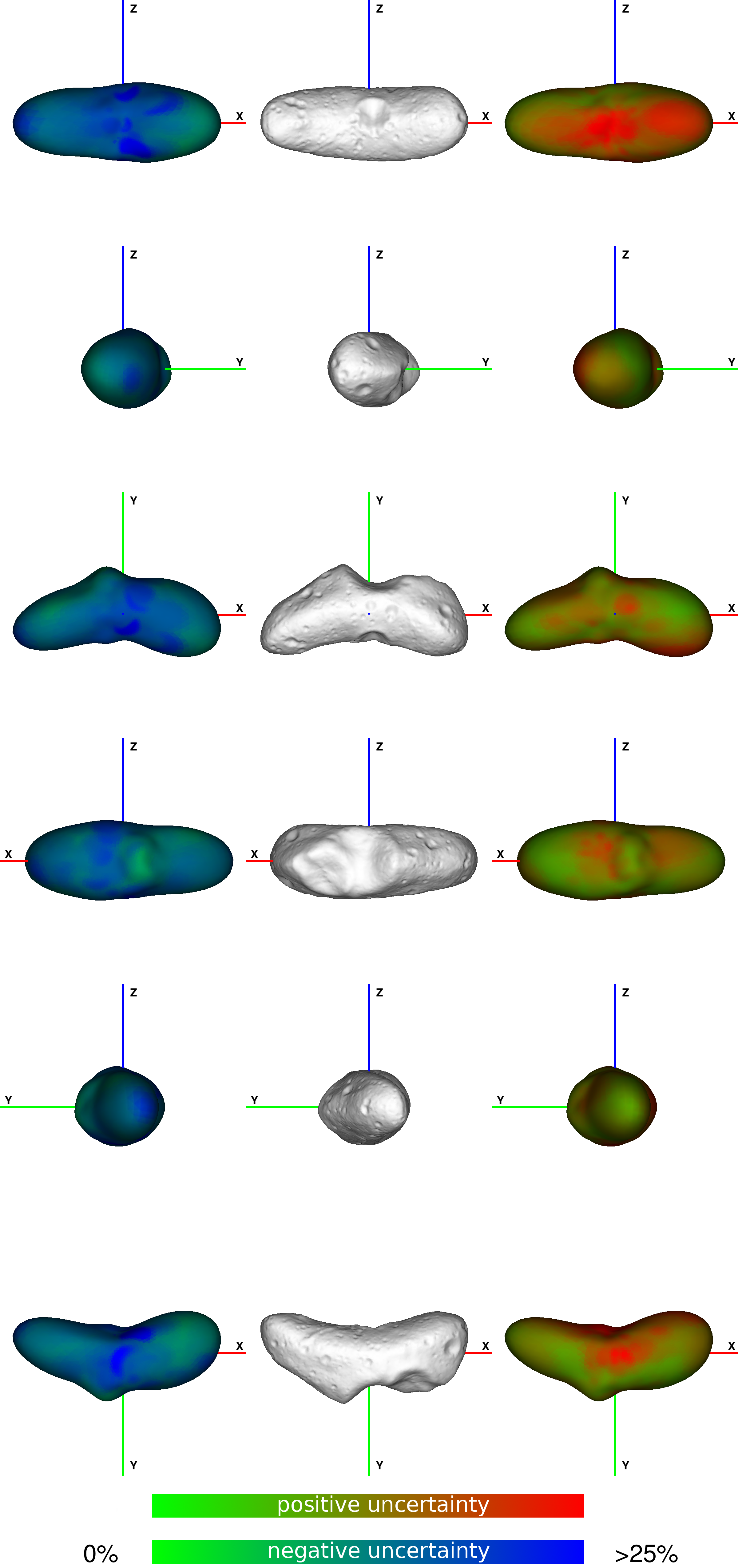}
	\caption{
	(433) Eros model \citep{SAGE} projections showing uncertainty values  (left
	and right columns) with comparison to model from NEAR Shoemaker mission
	 \citep{Zuber2000} (centre column).
	}
	\label{fig:eros_projections}
\end{figure}

The lightcurve-only non-convex model of Eros was modelled with SAGE method and
yielded great similarity to the real shape (see \cite{SAGE} for detailed
comparison of the shapes, and Fig.~\ref{fig:eros_projections} for comparison
with uncertainties). The lightcurve data was very rich (109 lightcurves) and
spanned over long period of 42 years (see Tab.~\ref{tab:eros_lc}).

(433) Eros and (243) Ida are similar cases, both real shapes being alike. Eros,
however, has spin axis orientation ($\lambda,\beta = 17\dd,9\dd$ ) much more
favorable in regard to lightcurve inversion compared to Ida's spin axis
($\lambda,\beta = 259\dd, -66\dd$).  The volume uncertainty for Eros is $36\%$,
much smaller in comparison to Ida's, partially due to more advantageous
geometry, and partially due to non-convex nature of the model. Uncertainties
show that general features, e.g. crater in the middle (which does not have round
edges otherwise preventing it from being represented in the lightcurves) or
global curvature of the model in $xy$ plane, are preserved, although the dent in
the middle (in $-y$ direction) could be either much deeper or shallower.

\subsection{(162173) Ryugu}
\input{img/tab_ryugu_obs}

\begin{figure*}
	\includegraphics[width=14cm]{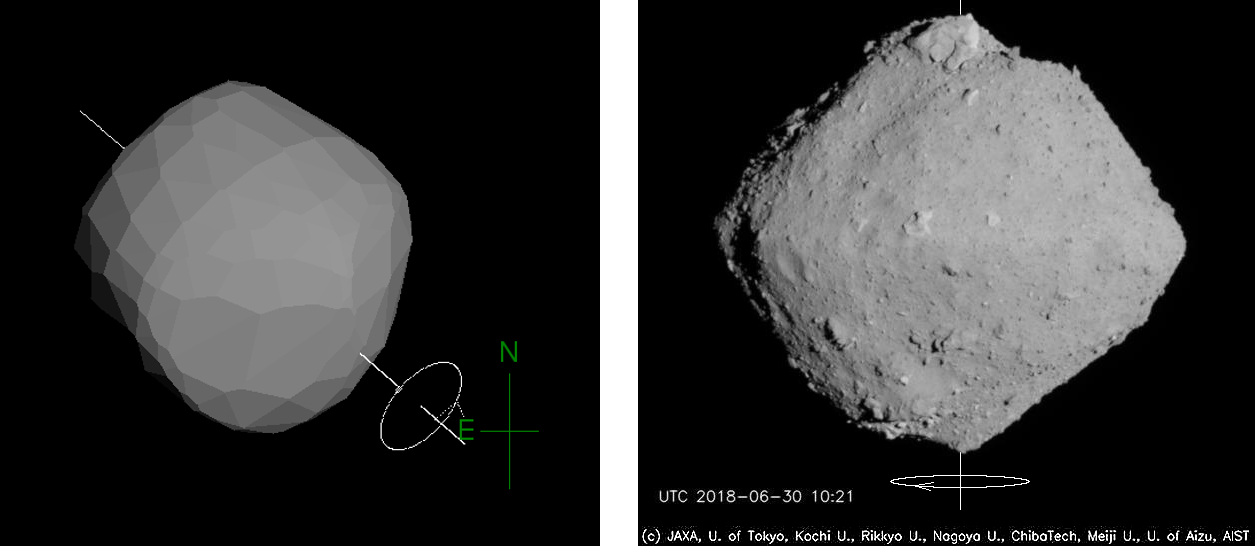}
	\caption{
		Comparison of asteroid (162173) Ryugu convex model
		\citep{Muller2017} (left) with image aquired by Hayabusa-2 mission
		(source: www.hayabusa2.jaxa.jp) on 30
	June 2018 (right).
	}
	\label{fig:ryugu_comparison}
\end{figure*}

Near Earth Asteroid (162173) Ryugu has been intensely studied since it has been
selected as Hayabusa-2 sample return mission target. The shape model used in
this work is taken from \cite{Muller2017}, where authors studied thermophysical
properties of Ryugu. The search for the spin axis and shape proved to be very
challenging and both optical and thermal data were used to find the solution,
since the lightcurve amplitude is very low (0.2 mag) with small signal-to-noise
ratio. See Tab.~\ref{tab:ryugu_lc} for the description of lightcurves for
this target. The shape was suspected to be nearly spherical with spin axis orientation
$\lambda, \beta = 340\dd, -40\dd$.  The Hayabusa-2 images revealed Ryugu's
bi-cone top shape and most probably rubble pile structure, as well as ecliptic
latitude of rotational axis orientation $\beta=-87.45 \pm 0.03\dd$
\citep{RyuguShapeDPS2018} (see Fig.~\ref{fig:ryugu_comparison} for comparison).

The model of Ryugu is far from what \textit{in situ} observations have unveiled,
particularly when it comes to spin axis orientation.
The uncertainties of $\lambda$ and $\beta$ justly reached the limits of scanning
area, which is $\pm30\dd$ from the nominal solution. The uncertainty of
rotational phase angle was as high as $190\dd$ with $10^{-3}$ rotational period
level of precision because of low lightcurve amplitude and, therefore, ability
to recognise distinct features in them.

Huge uncertainties in all tested parameters show that the method can serve as a
good indicator of model's robustness. Low quality observational data in case of
Ryugu produced large span of acceptable spin orientations (see Tab.~1 and Fig.~1
in \cite{Muller2017}). For the purpose of this work we used SAGE algorithm
\citep{SAGE} on Ryugu's data as well. The solution space was vast with plethora
of local minima with no shape and spin solution being significantly better than
other. Uncertainty assessment definitely revealed the problematic nature of
this target.

\subsection{z-scale from absolute photometry}

From the fact that absolute photometry was used during uncertainty assessment
procedure, the information on the best z-scale can be extracted. The results of
the search for the best z-scale are presented in the last column of
Tab.~\ref{tab:results}.

A problem not taken into consideration in the estimate presented in
Sec.~\ref{sec:absolute_limit} is the influence of the phase angle.  The nature
of the surface of the body has major influence on the slope parameter.
Additionally, the position of terminator coupled with the shape defines the
fraction of the body being in the shadow and subsequently the phase dependent
decrease in magnitudes.  Also, due to the rotation of a target and resulting
amplitude of reflected light, data points are scattered along the mean value.
The sampling is not random and can influence the value of slope parameter as
well. The z-scale is burdened with all if the effects combined.

\begin{figure}
	\includegraphics[width=8cm]{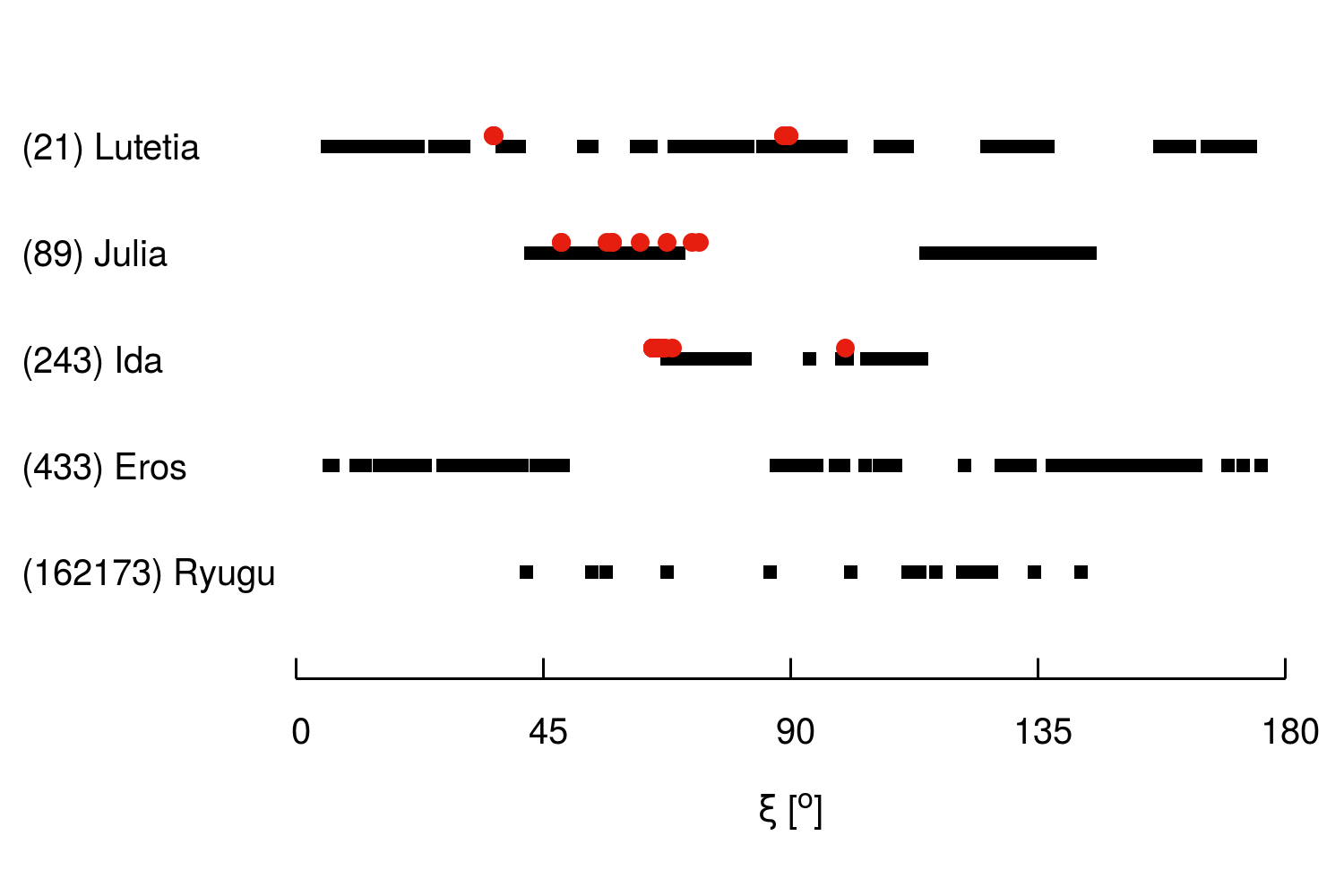}
	\caption{Plot showing absolute photometry aspect angles $\xi$ coverage. Black
		lines correspond to data from \protect\cite{Oszkiewicz2011} while red
		dots show Gaia DR2 measurements.}
	\label{fig:absolute_aspects}
\end{figure}

 The z-scales found correspond with the differences between lightcurve-based and
 \textit{in situ} models, particularly evident in case of (243) Ida. Moreover,
 the uncertainties of the z-scales correspond with the available ranges of
 aspect angles. Assuming nominal spin axes orientations, the absolute photometry
 aspect angles are presented on Fig.~\ref{fig:absolute_aspects}.

\section{Conclusions}


We created an uncertainty assessment method which can be applied to asteroid
models with reference to lightcurves and absolute sparse data in visual bands.
The method is effective as a measure of information content stored in
lightcurves -- something any lightcurve inversion method is lacking -- and has
informative role considering asteroid models' robustness adding new dimension
into the evaluation process. It was applied to a small sample of synthetic and
real targets showing that it can transform qualitative evaluation
(Sec.~\ref{sec:procedures}~and~\ref{sec:observations_limits}) into quantitative
one.

The results presented in this work indicate that shape and, therefore, other
physical parameter (e.g. volume or density) uncertainties of lightcurve-based
models are likely to be vastly understated. A large sample of models needs to be
examined in order to shed a new light on this matter.  By far, the unknown
extent of the model along its spin axis has the biggest influence on the volume
uncertainty. Establishing proper z-scale depends on the available aspect angles
and the photometric precision of the observational dataset. The lightcurve
inversion should definitely make use of absolute photometry, \hilight{especially
 precise and homogeneous Gaia dataset,} during the modelling process to
produce more reliable models volume-wise.

Preliminary assessment of volume uncertainty from available geometries, as shown
in Sec.~\ref{sec:observations_limits}, requires prior knowledge about the spin
axis. For the same reason, specific observing strategies could only be
established for, and applied to targets with known parameters in order to
improve the existing models. Collecting as much data as possible, evenly
distributed on the orbit and with the best achievable quality seems to be the
only general recipe one could give to assure low uncertainty of the models. The
Minor Planet Bulletin's continuously updated list of lightcurve photometry
opportunities (e.g. \cite{LcOpportunities2009}) created for observation
optimization concerning asteroid models can be utilised for exactly that
purpose.

When equivalent sphere diameters are being reported their uncertainties come
solely from the fit of the deterministic model (e.g. an ellipsoid, 3D shape from
lightcurve inversion) to the absolute measurements like stellar occultation
chords, adaptive optics images or thermal data. The uncertainties of the models
themselves are not considered at all. Making asteroid models' observation
predictions for various techniques (e.g.  lightcurves, stellar occultations,
adaptive optics images, radar delay-Doppler images, thermal emission) is
necessary to allow the computation of sizes taking models' uncertainty into
account. In order to achieve that, each data point has to have probability
distribution associated with it that has its source in model uncertainty.
Stochastic models of asteroids can be attained from the clone population created
during the uncertainty assessment procedure presented in this work. Exploiting
them will result in more realistic uncertainties of derived quantities.

Studying large sample of models, creating observation predictions and
incorporating other observational techniques into the uncertainty assessment
process are the areas which should definitely be explored further.

\section*{Acknowledgements}

The research leading to these results has received funding from the European
Union's Horizon 2020 Research and Innovation Programme, under Grant Agreement no
687378.

This work has made use of data from the European Space Agency (ESA) mission
{\it Gaia} (\url{https://www.cosmos.esa.int/gaia}), processed by the {\it Gaia}
Data Processing and Analysis Consortium (DPAC,
\url{https://www.cosmos.esa.int/web/gaia/dpac/consortium}). Funding for the DPAC
has been provided by national institutions, in particular the institutions
participating in the {\it Gaia} Multilateral Agreement.

%% file: img/tab_results.tex
\begin{table*}
\centering \caption{
	Compilation of results for models of (21) Lutetia, (89) Julia, (243) Ida,
	(433) Eros, and (162173) Ryugu. The uncertainties of volume $V$, rotational
	phase for reference epoch $\gamma_0$, rotational period $P$ and spin axis
	coordinates $\lambda$ and $\beta$ are reported. The last column describes
	the best value of z-scale calculated from absolute photometric data.
}
\label{tab:results}
\small\textsuperscript{1} \cite{Torppa2003};
\small\textsuperscript{2} this work;
\small\textsuperscript{3} \cite{Hanus2013b};
\small\textsuperscript{4} \cite{SAGE};
\small\textsuperscript{5} \cite{Muller2017};
\begin{tabular}{lcccccc}
\hline \hline
\textbf{model}      & $u(V)${[}\%{]}   & $u(\gamma_0)[\dd]$  & $u(P)${[}h{]} & $u(\lambda)[\dd]$     & $u(\beta)[\dd]$ 				& z-scale 				 	\\ \hline
(21) Lutetia$^1$    & $^{+9}_{-4}$     & $^{+4}_{-4}$        & $ 5 \times 10^{-6}$      & $^{+2}_{-1}$          & $^{+3}_{-3}$      & $1.08 ^{+0.01}_{-0.05} $           \\ \hline
(89) Julia$^2$      & $^{+19}_{-15}$   & $^{+2}_{-2}$        & $ 7 \times 10^{-6}$      & $^{+5}_{-2}$          & $^{+5}_{-3}$      & $0.92 ^{+0.21}_{-0.06} $           \\ \hline
(243) Ida$^3$       & $^{+51}_{-53}$   & $^{+12}_{-22}$      & $ 1 \times 10^{-5}$      & $^{+6}_{-4}$          & $^{+12}_{-22}$    & $0.52 ^{+0.5}_{-0.02} $           \\ \hline
(433) Eros$^4$      & $^{+14}_{-20}$   & $^{+10}_{-4}$       & $ 3 \times 10^{-6}$      & $^{+2}_{-2}$          & $^{+1}_{-7}$      & $1.02 ^{+0.09}_{-0.09} $           \\ \hline
(162173) Ryugu$^5$  & $^{+52}_{-59}$   & $^{+100}_{-90}$     & $ 1 \times 10^{-3}$      & $^{>+30}_{<-30}$      & $^{>+30}_{<-30}$ 	& $0.92 ^{+0.09}_{-0.11} $           \\ \hline
\end{tabular}
\end{table*}


%% file: img/tab_lutetia_obs.tex
\begin{table*}
	\caption{Details of the lightcurve data used for (21) Lutetia modelling and
		uncertainty assessment.  N$_{lc}$ -- number of lightcurves per
		apparition,  $\alpha$
 -- phase angle, $\lambda$ -- ecliptic longitude, $\beta$ -- ecliptic latitude.}
\label{tab:lutetia_lc}
\begin{tabular}{ccccccc}
\hline
\hline
  Apparition & Year & N$_{lc}$ & $\alpha$ $[^{\circ}]$ & $\lambda$ $[^{\circ}]$
& $\beta$ $[^{\circ}]$ & reference\\
\hline
  1 & 1962 & 30 &  28     & 18        & -3 & \cite{ChangAndChang1963} \\
  2 & 1981 & 76 & 6 -- 21 & 347 -- 14 & -3 & \cite{Lupishko1983} , \cite{Zappala1984} \\
  3 & 1983 & 4 & 3 -- 28 & 130 & 2 & \cite{Zappala1984}, \cite{Lupishko1983} \\
  4 & 1985 & 75 & 5 -- 24 & 34 -- 46 & -2 & \cite{Dotto1992}, \cite{Lupishko1987b} \\
&&&&&& \cite{Lagerkvist1995b} \\
 5 & 1986 & 1 & 7 & 60 & -1 &  \cite{Lupishko1987b}\\
 6 & 1991 & 4& 16 & 173 & 3 &    \cite{Lagerkvist1995b} \\
 7 & 1995 & 12 & 2 & 178 & 3 & \cite{Denchev1998}\\
 8 & 1998 & 6 & 26 & 115 -- 120 & 2 & Denchev (2000), private communication\\
 9 & 2003 & 20 & 5 -- 29 & 223 -- 229 & 2 & \cite{Carry2010} \\
 10& 2004 & 7& 17 & 5 & -3 &   \cite{Carry2010} \\
 11 & 2005/2006 & 14 & 12 -- 21 & 134 -- 142 & 3 & \cite{Carry2010} \\
 12& 2007 & 1& 3 &  212 &  2& \cite{Carry2010} \\
 13& 2008/2009 & 23 & 4 -- 25 & 69 -- 95 & -1 -- 1 &  \cite{Carry2010} \\
14 & 2010 & 12 & 7 -- 16 & 165 & 2 & \cite{Carry2010} \\
\hline
\end{tabular}
\end{table*}

%% file: img/tab_julia_obs.tex
\begin{table*}
	\caption{Details of the lightcurve data used for (89) Julia modelling and
		uncertainty assessment.  N$_{lc}$ -- number of lightcurves per
		apparition,  $\alpha$
 -- phase angle, $\lambda$ -- ecliptic longitude, $\beta$ -- ecliptic latitude.}
\label{tab:julia_lc}
\begin{tabular}{ccccccc}
\hline
\hline
  Apparition & Year & N$_{lc}$ & $\alpha$ $[^{\circ}]$ & $\lambda$ $[^{\circ}]$
& $\beta$ $[^{\circ}]$ & reference\\
\hline
  1  & 1968 &  5  &     5    &  326 -- 329  & 4   & \cite{Vesely1985}\\
  2  & 1972 &  8  &     5 -- 13    &  318 -- 322  & 2   & \cite{Schober1975}\\
  3  & 2009 &  18  &     19 -- 24    &  0 -- 6 & 13   & \cite{Hanus2013}\\
  4  & 2017 &   7  &     18 -- 21    &  331 -- 334 & 6   & \cite{Warner2018}\\
\hline
\end{tabular}
\end{table*}

%% file: img/tab_ida_obs.tex
\begin{table*}
	\caption{Details of the lightcurve data used for (243) Ida modelling and
		uncertainty assessment.  N$_{lc}$ -- number of lightcurves per
		apparition,  $\alpha$
 -- phase angle, $\lambda$ -- ecliptic longitude, $\beta$ -- ecliptic latitude.}
\label{tab:ida_lc}
\begin{tabular}{ccccccc}
\hline
\hline
  Apparition & Year & N$_{lc}$ & $\alpha$ $[^{\circ}]$ & $\lambda$ $[^{\circ}]$
& $\beta$ $[^{\circ}]$ & reference\\
\hline
  1 & 1980 &   9 &  16 & 321 & 0 &   			  \cite{Binzel1993}\\
  2 & 1984 &   6 &   3 & 244 & -1 &   			  \cite{Binzel1987}\\
  3 & 1988 &   15 &  23 -- 15 & 158 -- 180 & 0 &  \cite{Binzel1993}\\
  4 & 1990 &  130 &  18 & 339 -- 5 &  1 &   	  \cite{GonanoBeurer1992}, \cite{Binzel1993}\\
  5 & 1991/1992 &  67 &  2 -- 30 & 67 -- 93 & 1 & \cite{Binzel1993}, \cite{Mottola1994}\\
 6 & 1992/1993 & 79  & 12 -- 29 & 165 -- 182 & -1 & \cite{Binzel1993}, \cite{Mottola1994},\\
									       &&&&&& \cite{SlivanBinzel1996}\\
 7 & 1993 &   3 &  25 &  76 &  1 &                \cite{Binzel1993}\\
\hline
\end{tabular}
\end{table*}

%% file: img/tab_eros_obs.tex
\begin{table*}
	\caption{Details of the lightcurve data used for (433) Eros modelling and
		uncuncertainty assessment. N$_{lc}$ -- number of lightcurves per
		apparition,  $\alpha$
 -- phase angle, $\lambda$ -- ecliptic longitude, $\beta$ -- ecliptic latitude.
	}
	   \label{tab:eros_lc}
	 \begin{tabular}{ccccccc}
		 \hline
		\hline
		   Apparition & Year & N$_{lc}$ & $\alpha$ $[^{\circ}]$ & $\lambda$ $[^{\circ}]$ & $\beta$ $[^{\circ}]$ & reference\\
		\hline
		   1  & 1951/1952 &  28  & 19 -- 59 & 5 -- 119 & -10 -- 22 & \cite{beyer53},\\
		   2  &   1972    &   1  &    17      & 342      &     9       & \cite{dunlap76}\\
		   3  & 1974/1975 &  68  &  9 -- 44 & 53 -- 158 & -31 -- 33 & \cite{cristescu76}, \cite{dunlap76},\\
		&&&&&& \cite{millis76},\\
		&&&&&& \cite{miner76}, \cite{pop76},\\
		&&&&&& \cite{scaltriti76}, \cite{tedesco76}\\
		  4   & 1981/1982 &  4   & 29 -- 54 & 42 -- 126 & -17 -- 37 & \cite{drummond85}, \cite{harris99}\\
		  5   &   1993    &  8   & 1 -- 18 & 296 -- 308& -1 -- 4  & \cite{krugly99}\\
		\hline
	 \end{tabular}
\end{table*}

%% file: img/tab_ryugu_obs.tex
\begin{table*}
	\caption{Details of the lightcurve data used for (162173) Ryugu modelling and
		uncertainty assessment.  N$_{lc}$ -- number of lightcurves per
		apparition,  $\alpha$
 -- phase angle, $\lambda$ -- ecliptic longitude, $\beta$ -- ecliptic latitude.}
\label{tab:ryugu_lc}
\begin{tabular}{ccccccc}
\hline
\hline
  Apparition & Year & N$_{lc}$ & $\alpha$ $[^{\circ}]$ & $\lambda$ $[^{\circ}]$
& $\beta$ $[^{\circ}]$ & reference\\
\hline
1 & 2007 & 23 & 41 -- 25 & 307 -- 350 & 5 &
\multirow{9}{*}{\begin{tabular}[x]{@{}c@{}} \cite{Muller2011,Muller2017},\\ T.~M{\"u}ller, private communication \end{tabular} } \\
2 & 2007 & 12 & 49 -- 79 & 13 -- 61 & 5 -- 1 &\\
3 & 2008 & 11 & 88 -- 54 & 127 -- 186 & -5 &\\
4 & 2011 & 1 & 55 & 349 & 6 &\\
5 & 2011 & 1 & 76 & 62 & 1 &\\
6 & 2012 & 34 & 0 -- 49 & 215 -- 277 & -3 -- 3 &\\
7 & 2013 & 2 & 47 & 57 & 1 &\\
8 & 2013 & 2 & 52 & 146 & -6 &\\
9 & 2016 & 12 & 43 -- 17 & 264 -- 305 & 1 -- 5 &\\
\hline
\end{tabular}
\end{table*}